# Mixing in Three-dimensional Cavity by Moving Cavity Walls

(*)Alex Povitsky

Email: povitsky@uakron.edu

Department of Mechanical Engineering

The University of Akron, Akron OH USA 44325-3903

**Abstract**

The mixing in three-dimensional enclosures is investigated numerically using flow in cubical cavity as a geometrically simple model of various natural and engineering flows. The mixing rate is evaluated for up to the value of Reynolds number Re=2000 for several representative scenarios of moving cavity walls: perpendicular motion of the parallel cavity walls (Case A), motion of a wall in its plane along its diagonal (Case B1), motion of two perpendicular walls outward the common edge (Case B2), and the parallel cavity walls in motion either in parallel directions (Case B3) or in opposite directions (Case B4). The mixing rates are compared to the well-known benchmark case in which one cavity wall moves along its edge (Case C). The intensity of mixing for the considered cases was evaluated for (i) mixing in developing cavity flow initially at rest, which is started by the impulsive motion of cavity wall(s), and (ii) mixing in the developed cavity flow. For both cases, the initial interface of the two mixing fluids is a horizontal plane located at the middle of the cavity. The mixing rates are ranked from fastest to slowest for twenty time units of flow mixing. The pure convection mixing is modeled as a limit case to reveal convective mechanism of mixing. Mixing of fluids with different densities is modeled to show the advantage in terms of mixing rate of genuinely 3-D cases A and B1. Grid convergence study and comparison with published numerical solutions for 3-D and 2-D cavity flows are presented. The effects of three-dimensionality of cavity flow on the mixing rate are discussed.

**Introduction**

Because of the variety of natural, industrial and biomedical prototype applications, steady-state 2-D cavity flows have been widely studied by both experimental and numerical investigations, see reviews[1,2]. Studies of 3-D cavity flows started after the pioneering experimental work[3]. However, very few studies have been conducted on the transient flow establishment and mixing phase[4,5]. Ref[6] quantifies the mixing characteristics of a two-dimensional, lid-driven blinking Stokes flow ($Re<1$) to evaluate fluidic components that are critical parts of micro- and nano-scale systems. The latter can be used for detecting both chemical and biological agents and explosives, monitoring the environment for hazardous chemicals or toxins, and diagnosing and treating medical problems. These fluidic components can be used for transporting and mixing small amounts of materials that are subsequently analyzed or delivered to predetermined sites. The above listed applications include fluid dynamics in small channels that have etched or engraved geometric features, such as grooves[6].

Most results to date of chaotic advection (generation of small-scale structures by the stretching and folding in fluid without turbulence) have been found by examination of Lagrangian structures in 2-D low-Re flows. These structures between the generically integrable states in the Stokes (Re→0) and Euler (Re→∞) limits remain unexplored to date (see recent review[7]). In 3-D flows there is an explosion of complexity in the number of possible Lagrangian structures and connections between them compared to their counterparts in 2-D. The 3-D equations of motion lack the well-defined Hamiltonian structure of 2-D configurations (see[7] and references therein).

Fluid stretching and folding increases the interface(s) across which diffusion occurs, thereby increasing the mixing rate[8]. Rapid mixing through stirring by moving boundaries depends on the dynamics of the fluid. However, the mixing in enclosures literature has focused primarily on Stokes' flow models (see[8] and references therein).

The goal of the current study is to quantify and compare mixing rates in 3-D cubical cavity for developed and developing laminar flowfield at Re of the order of $10^3$ for several representative cases of moving cavity walls for equal strength of diffusion and convection for mixing and for limit cases of diffusion- and convection- dominated mixing. By Ref[1], recirculating cavity flows generated by the motion of one or more of the containing walls are not only technologically important, but they are also of great scientific interest because they display almost all fluid mechanical phenomena in the simplest of geometrical settings. The analytical Stokes' solution[8] is only correct for Re→0. To include the effects of inertia in cavity flow, the governing Navier-Stokes equations should be solved computationally for a finite Reynolds number.

By computational results obtained in the current study, some set-ups of moving walls for cavity flow exhibit relatively slow mixing compared to other configurations that mix well. Each case introduces its degree of three-dimensionality ranking from nearly 2-D flow (with the exception of end walls) to essentially 3-D flow with enhanced mixing.

The normalized variance of concentration is appropriate to evaluate mixing (see[5-9] and references therein). From an initially unmixed state, the variance of concentration decreases over time, indicating that mixing has occurred to a required degree. The current study compares cubical cavity flows configurations caused by motion of cavity wall(s) (see Fig. 1) in order to evaluate the dynamics of variance of concentration and to explain physical reasons for delay in mixing in certain areas of flowfield. In particular, it will be shown in the current study that the 3-D flow in Case A (Fig. 1a) has faster mixing compared to other above listed cases. This flow, which has recently been proposed by the author[9], being driven by the perpendicular motion of opposite walls also referred to as top and bottom cavity walls in the current study to follow literature terminology. For Case A, the axes of primarily vortices are in the $x$ and $z$ perpendicular directions, which makes the flowfield three-dimensional. A limit case with the Schmidt number (Sc) tending to infinity while keeping Re=2000 is considered in order to evaluate the role of convection mixing separately of mass diffusion. The Sc is a dimensionless number defined as the ratio of momentum diffusivity and mass diffusivity. The formation of larger contact surface of mixing fluids by convection in Case A compared to the benchmark Case C is shown. The practical value of considering Sc>>1 is that for liquid-liquid systems values of Sc are typically of the order of 1000[10].

The cubic cavity flow in which the top wall moves in its plane along its diagonal was introduced by the author[11] (Case B1 in Fig.1b). The prototype flows are those typical for urban air pollution[12] and pneumatic transport of powder materials[13]. Refs[14,15] propose the 3-D cavity flow in which the top wall moves to the right, while the left vertical wall moves down with the same constant velocity (see Fig.1c, case B2). For Case B2, the flows recirculate in upper and lower cavity prisms, which are separated by the cavity diagonal plane that forms the plane of symmetry.

The prior study[16] modeled flows in 2-D cavities in which the top and bottom walls move either in the same direction or in opposite directions with the same speed. Different shapes and sizes of streamline patterns were obtained[16] for various values of Reynolds numbers and cavity aspect ratios. The 3-D extension of this motion is introduced[17]. In the current study, the mixing pattern is evaluated for 3-D cavity with its top and bottom moving. For Case B3 (Fig. 1d), the top and bottom walls move in a parallel fashion. For Case B4 (Fig. 1e), the top and bottom walls are in reverse motion.

For Case C (Fig. 1f), the flow inside the cavity is generated by the translation of one cavity wall along its edge. This particular configuration[3] is widely used for validation of numerical methodologies[18]. It should be noted that cases B2-B4 and C can be reduced to their 2-D analogues, while cases A[9] and B1[11] are essentially three-dimensional, and do not have their 2-D counterparts. Nevertheless, the current study shows that three-dimensional effects on mixing are significant for cases B2-B4 and C.

To evaluate the time interval for cavity flow development[6], the developing flowfield was computed for $T \leq 20$, in which time T is normalized by the ratio of the cavity size to moving lid speed, $L/U$. At T=0 the boundary(ies) start moving impulsively as required by each particular case (see Fig. 1). By physical experiments[4], the cavity flowfield was recorded at regular dimensionless time steps: $\Delta T=1$, for $0 \leq T \leq 12$. By Ref[4], cavity flowfield reached a quasi-steady state by $T=8-12$, which was measured by the stagnation of vortex-core positions, secondary eddy sizes, and velocity profiles. In the current study, the temporal derivative of concentration deviation is introduced to evaluate the mixing rate. It is shown that the maximum of mixing rate corresponds to T<20 for all considered cases and toward T=20 the mixing rate slows down.

Vorticity magnitude and velocity vector field are presented in order to identify fluid structures appearing in the developed cavity flow for above mentioned mixing cases and to discuss their effect on mixing rate.

The study can be extended to non-rectangular lid-driven cavities and two-phase flows[19,20]. In the future research, the cavity flow mixing studies can be extended to non-Newtonian fluids. The authors[21] investigated the circulating flow of power-law fluids inside a square cavity of height H for both the parallel and reverse motions of two facing lids. During the parallel motion of the lids, there are two counter-rotating primary vortices, and the streamlines in one half of the cavity is the mirror image of the other with respect to the line y/H = 0.5. During anti-parallel wall motion, a single primary vortex develops in the cavity. In the future, the current comparative evaluation of intensity of mixing for various 3-D cavity flows can be extended to non-Newtonian flows typical for polymer engineering, food processing, dynamics of drilling fluid, and the manufacture of energy materials.

The study is composed as follows: the mathematical description of fluid motion in 3-D cavity, numerical methodology for the solution of the mathematical model, grid refinement study and methodology for computing the dynamics of variance of concentration are given in Section 2. Computational results of the integral evaluation of mixing rate, local features of flowfield, and mixing in the middle vertical section for each case are presented in Section 3. In Section 4, the 3-D effects on flowfield and mixing, including finiteness in the span-wise direction for each case, are discussed. The mixing in limit cases of small and large Schmidt numbers corresponding to diffusion- and convection- dominated mixing are considered in Section 5. The mixing of fluids with different densities is presented in Section 6. Conclusions are presented in the last section.

**2. Mathematical model, validation of computational approach and evaluation of mixing rate**

The governing equations in all sections except Section 6 are the three-dimensional incompressible transient continuity and Navier-Stokes equations in Cartesian coordinates *(x, y, z)* describing conservation of total mass, momentum and the mass of the first fluid:

$$\frac{\partial u}{\partial x} + \frac{\partial v}{\partial y} + \frac{\partial w}{\partial z} = 0,$$

(1)

$$\frac{\partial u}{\partial t} + u\frac{\partial u}{\partial x} + v\frac{\partial u}{\partial y} + w\frac{\partial u}{\partial z} + \frac{\partial p}{\partial x} - \frac{1}{Re}\nabla^2 u = 0,$$

$$\frac{\partial v}{\partial t} + u\frac{\partial v}{\partial x} + v\frac{\partial v}{\partial y} + w\frac{\partial v}{\partial z} + \frac{\partial p}{\partial y} - \frac{1}{Re}\nabla^2 v = 0,$$

$$\frac{\partial w}{\partial t} + u\frac{\partial w}{\partial x} + v\frac{\partial w}{\partial y} + w\frac{\partial w}{\partial z} + \frac{\partial p}{\partial z} - \frac{1}{Re}\nabla^2 w = 0,$$

(2)

$$\frac{\partial c}{\partial t} + u\frac{\partial c}{\partial x} + v\frac{\partial c}{\partial y} + w\frac{\partial c}{\partial z} - \frac{1}{Sc\,Re}\nabla^2 c = 0,$$

(3)

where *u, v,* and *w* are the components of the velocity in the x, y and z directions, respectively, Sc=ν/D is the Schmidt number representing the ratio of kinematic viscosity, ν=μ/ρ, to mass diffusivity, D. Laplace operator $\nabla^2 F = \frac{\partial^2 F}{\partial x^2} + \frac{\partial^2 F}{\partial y^2} + \frac{\partial^2 F}{\partial z^2}$ and *c(x,y,z,t)* is the mass concentration of the first fluid in Eq. (3). The governing variables and coordinates are normalized by the moving wall speed *U,* length of cube edge *L* and fluid density *ρ*. Gravity force is neglected in Eqs. (2). The surface tension between two fluids is neglected in Eqs. (2), that is, the Weber number is assumed to be much larger than unity.

It should be noted that in Sections 3 and 4 both mixing components have the value of Schmidt number *Sc=1*, therefore, the intensity of mass and momentum diffusion are the same. The limit cases of Sc→∞ and Sc→0 are considered in Section 5. In Sections 3-5 the flow is incompressible as both fluids have the same density and satisfy the incompressible form of continuity and momentum equations, Eqs. (1-3). The compressible equations for mixing of heavy and light fluids are introduced and solved in Section 6.

The boundary conditions are the no-slip and non-penetrating conditions at the stationary and moving walls. Solutions of the incompressible Navier-Stokes system of partial differential equations with appropriate boundary conditions depend on a single parameter: the Reynolds number, $Re=\rho UL/\mu$, where $\mu$ is viscosity of fluid respectively. For normalized variables used in the current study $U, L$ and $\rho$ are taken equal to unity so as $Re=1/\mu$.

To evaluate mixing quantitatively, the cavity is filled with two fluids having the same viscosity; therefore, the flow remains incompressible and controlled by a single parameter: the Reynolds number. The first fluid occupies the upper half of cavity (y>0.5), while the second fluid fills the lower half of cavity (y<0.5). Thus, the plane y=0.5 divides two species with the same properties. The variance of concentration, $\sigma^2$, is the degree of non-mixing of these two fluids. The variance of concentration is quantified by the mean-squared variability

$$\sigma^2 = (\int (c - 0.5)^2 \, dV)/V, \quad (4)$$

where $dV$ is a cavity volume differential element, and V is the cavity volume, which is equal to unity in the current study. At t=0, initial value of $\sigma^2$ is equal to $\sigma_0^2 = 0.25$ for all cases because the value of integrand in above equation is equal to *(1-0.5)²=0.5²=0.25* for the upper half of cavity and *(0-0.5)²=0.5²=0.25* for the lower half of cavity. When two flow components mix well with each other, the variable $\sigma^2$ tends to zero. The deviation of concentration, $\sigma$ is the square root of the variance of concentration. The deviation of concentration is another related measure of how spread out the concentration is, which is used in the current study.

The value of normalized deviation of concentration, $\sqrt{\frac{\sigma^2}{\sigma_0^2}}$, is plot in Figs. 2-3 as a function of unit-less time $T=tU/L$ for mixing of two fluids. The temporal rate of change of normalized deviation of concentration, $d(\sigma/\sigma_o)/dt$, shows the average-over-cavity mixing rate and will be compared for considered cases in the current study (Fig. 4).

The 3-D cavity flowfield is obtained by the numerical solution of the three-dimensional viscous fluid flow equations (1-2), as described in the prior study of the author[9] by using ANSYS/Fluent finite-volume software with second-order upwind schemes for convective terms[22,23] and second-order central scheme for viscous terms. ANSYS/Fluent software uses the Semi-implicit Method for Pressure-linked Equations (SIMPLE)[24] to resolve velocity and pressure coupling in non-linear Navier-Stokes equations (Eqs. (2)). Comparison of the author's computations by using ANSYS/Fluent with prior experimental[3] and computational results[25,26,27] for the benchmark Case C appears in Ref[9], Section 2, Figure 3. In Ref[6], the grid convergence study for Cases A and C is presented. To verify the accuracy of solutions for the two-dimensional and three-dimensional single lid-driven cavity flows, comparison with results obtained by other numerical methods is presented in Appendix.

The implicit second-order temporal discretization method[22] is used with the time step $\Delta t=0.01$.

To evaluate grid-independence of the intensity of mixing, the deviation of concentration for Case B1 was evaluated first for developing cavity flow initially at rest, *(u,v,w)=0* at *t=0*. Two uniform numerical grids with grid steps h=0.005 (201 x 201 x 201 grid) and h=0.01 (101 x 101 x 101 grid) are used. The integral (4) was computed for the first twenty units of normalized time for Re=2000 and 1000. The computed value of $\sigma/\sigma_0$ is shown in Fig. 2a. Results obtained by these two grids appear to be close to each other.

The second cavity flow set-up serves the purpose of evaluation of mixing of fluid in the developed cavity flow starting at *T=20*, see Fig. 2b. As opposed to the prior situation, the mixing in this case occurs in fully-developed cavity flowfield. The fully-developed flowfield for *Re=2000* and *1000* was computed as described above. The initial location of interface of two fluids is at *y=0.5* as it was in the prior case. The integral (4) was computed for Case B1 for the first twenty units of normalized time after set-up of the interface of two fluids (*20≤T≤40*) as depicted in Fig. 2b.

Present computational results show that the concentration variance as a function of time practically coincide for these grids; therefore, the *201 x 201 x 201* grid is selected for computations presented in the next sections. Studies[28,29] show that the 3-D cavity flow (Case C) becomes unstable (transit to turbulence) when *Re>2000*. The oscillatory instability is observed for the critical value of Reynolds number[30] $Re_{cr}$ = 2320, therefore, for Re≤2000 the solutions do not bifurcate. To limit the current study to stable flows, *Re=2000* is selected for the evaluation of mixing in the next sections.

Figure 2c show the effect of the magnitude of computational time step on the variation of concentration. The time steps from $\Delta t=0.08$ to $\Delta t=0.01$ were used and the deviation of concentration as a function of time nearly coincide for $\Delta t=0.02$ and $\Delta t=0.01$ that confirms the selection of time step $\Delta t=0.01$.

To estimate the diffusion error of the numerical scheme used, the lead coefficient in modified (equivalent) differential equation is evaluated (see[31], p. 120). Numerical viscosity for transient advection equation discretized with the second-order upwind scheme is $O(\frac{c\Delta x^2}{6})$, where c is the local advection velocity. As $c \leq 1$ for the normalized velocity in cavity, $\frac{c\Delta x^2}{6} < \frac{0.005^2}{6} < 10^{-5}$ for $201^3$ grid.

Numerical viscosity for transient diffusion equation is $O(-0.5\alpha^2\Delta t + \frac{\alpha\Delta x^2}{12})$, where $\alpha=1/Re$ is the local diffusion coefficient (see[31], p. 127). To evaluate components of the numerical viscosity, $-0.5\alpha^2\Delta t = -0.5\left(\frac{1}{2000}\right)^2 0.01 < 10^{-8}$ and $\frac{\alpha\Delta x^2}{12} = \frac{0.005^2}{2000 \times 12} < 10^{-8}$. Therefore, the level of numerical viscosity for both advection and diffusion is orders of magnitude smaller than the physical diffusion, which is equal to $1/Re=1/2000 \sim 10^{-3}$.

3. Computational results

In this section the mixing results obtained by the computational model are discussed for developing flow and developed flow with $Sc=1$ while pure convection mixing is discussed in Section 5. For developing flow isolines of concentration at $T=5$ are shown for Cases A, C, B1-B4 in Figs 5a-10a, respectively. For developed flow isolines of concentration (Figs. 5b-10b), vorticity (Figs. 5c-10c) and velocity vector field (Figs 5d-10d) are shown at *T=20*.

For the benchmark Case C it was shown[33] that the central vortex rotates as a rigid body, with constant vorticity level, with the vortex center located near the center of cavity for Re>1000. Within a vortex rotating as a rigid body, the exchange of momentum by shear stress is zero as there is no deformation of fluid particles. Nevertheless, the vortex can cause mixing of two fluids by convection. The vorticity for considered cases is shown to identify constant-vorticity areas of rigid body rotation and elevated vorticity areas with significant shear stress. The vorticity magnitude interval $0 \leq |\omega| \leq 5$ is taken in all figures to follow[33] while larger vorticity levels including those near the cavity walls are not depicted.

For developing cavity flow, the mixing rate is slow for the first five time units. The normalized deviation of concentration remains close to unity and nearly the same for all considered cases, except for B2 (Fig 3a). The rate of mixing reaches its minimum when $T<5$ for all cases but B2, see Fig. 4a. For the latter case, the mixing starts early, as the moving left vertical cavity wall is located next to the interface of fluids at y=0.5 (see Fig 8a). For all other cases, the fluids in the cavity are still separated at T=5 (see isolines of concentration for Cases A, C, B1, B3, B4 in Figs. 5a, 6a, 7a, 9a and 10a). The interface between two fluids has been deformed compared to its initial horizontal location at y=0.5 but has not been ruptured yet to enhance the convective mixing. The most prominent deformation of interface, which allow larger interface surface and, consequently, more intense diffusion is shown for Case B4 (Fig. 10a) followed by Case A (Fig. 5a). Case C (Fig. 6a) has deformation of interface to a lesser degree compared to B4 and A.

For Case B1 (Fig. 7a), the degree of deformation of interface varies with its location. Case B1 can be viewed as a set of 2-D cavities with varying length-to-depth ratio, from relatively shallow diagonal section with its length to depth ratio $\sqrt{2} : 1$ to deeper peripheral sections. The flowfield within these cavities is connected by the flow in planes perpendicular to them. For the diagonal section the deformed interface is significantly curved and reach the cavity bottom (right cavity in Fig. 7a). For the diagonal plane and for the plane parallel to diagonal plane with the 1: 1 length to depth ratio (central cavity in Fig. 7a), the interface begins to rupture and allow convective mixing. On the contrary, for the peripheral plain parallel to diagonal plain with the length to depth ratio 0.5 : 1 (left cavity in Fig. 7a) the interface is disturbed to a lesser degree.

For Case B2 (Fig. 8a), the fluid located at the upper part of cavity is dragged by the left vertical wall moving down and mixes with the second fluid by vortex forming in the lower left part of cavity. For Case B3 (Fig. 9a) the interface remains flat at $y=0.5$ because of the symmetry of the set-up. The diffusion of two fluids across the interface and spread of fluids by vertical motion within upper and lower cavity halves leads to the moderate smearing of the interface.

For T>5, the deviation of concentration and rate of mixing are different among considered cases (see Fig. 3a and Fig. 4a). The time rate of deviation of concentration (Fig. 4a) reaches its maximum between 10 and 15 time units. Afterwards the rate decreases for all cases that justifies the simulation time of 20 time units.

For Case A, not only the maximum of temporal derivative is bigger than that for other cases but also the time interval at which the temporal derivative is close to its maximum, *8<T<14,* is substantially larger than that for other considered cases (Fig. 4a). Cases B4 and A have similar mixing rates till T≈12, which are substantially faster compared to other cases. Case A is substantially three-dimensional because the top and bottom walls of the cavity move in perpendicular directions[9]. Case B4, in which the top and bottom walls of cavity move in opposite directions, forms a bigger primary vortex with strong shear stress at bottom wall (Fig. 10c,d) compared to the benchmark Case C (Fig. 6c,d), in which only the top cavity wall moves. For T>12, Case B4 slows down in terms of completeness of mixing (Fig. 3a) and the temporal rate becomes smaller than that for Case A (Fig. 4a).

The concentration field at T=20 is significantly more divided into pieces by stretching and folding compared to Case C; this leads to formation of smaller size areas of non-mixed fluids with a well-developed surface (compare Fig. 5b to Fig. 6b). For Case C, the big and relatively unmixed area rotates by the rigid body rotating vortex (red area in Fig. 6b), therefore, the mixing rate for Case A is larger than that for the benchmark case C. The vorticity for Case A has areas of larger magnitude in the central part of the cavity (compare Figs. 5c to 6c) and the central rigid body rotating vortex is much smaller in size compared to Case C (compare Fig. 5d to Fig 6d). The areas of up-down motion surrounding the central vortex in Case A (Fig. 5d) lead to faster mixing.

For Case B1, the mixing rate is somewhat faster compared to B3 and C; however, the deviation of concentration becomes equal to that for Case C at *T≈14* time units and case C has faster mixing rate compared to B1 for *T>11*. For Case B1, the mixing at the mean diagonal plane is more intense compared to that for Case C (compare Figs. 7b to 6b). The large unmixed area in Case C (Fig. 6b) no longer exists in Case B1 (Fig. 7b). The rigidly rotating vortex in the central diagonal plane is shifted toward the upper lid and forward compared to benchmark Case C with formation of up-down flow in the front part of the central diagonal plane (Fig. 7d) and moderate vorticity zone protruding from the upper right corner toward the lower left corner along the second diagonal of the central diagonal plane (Fig. 7c). Big areas of nearly complete mixing are formed in the central diagonal plane (green color in Fig. 7b), however, the back and near-bottom parts of the off-diagonal sections (Fig. 7b) have areas of unmixed fluids.

For Case B2 (Fig. 8), the two fluids are initially separated by interface at $y=0.5$, while at later time moments the diagonal plane of symmetry is created in a way that separates two fluids and slows down the mixing (Fig. 8b). Formed rigid body rotating vortices in upper right and lower left corners rotate fluid within each half-cavity prism separately (Fig. 8c,d). At earlier time moments (Fig. 8a), the transition of the interface between mixing fluids, from the horizontal to diagonal, creates faster mixing. As a result, the mixing rate for Case B2 is the fastest at earlier times (T<5), and slows down as the time progresses (see Figs. 3a and 4a).

For Case B3 (Fig. 9), the flowfield symmetry plane at $y=0.5$ coincides with the interface of two mixing fluids. There is no flow across the symmetry line; therefore, there is no convection of species across the interface. The formed symmetric rigid body rotating vortices (Fig. 9c,d) assist in mixing the fluids, which were diffused through the interface, within upper and lower halves of the cavity. However, the big unmixed volumes of fluids are formed in central parts of the halves of the cavity. Slow mixing for Case B3 is limited mainly by diffusion across the symmetry line, with the exception of 3-D effects in the plane perpendicular to the central vertical cross-section (as described in the next section).

Case B4 (Fig. 10) has the fastest mixing rate of the four cases listed above. Similar to Case C, the prominent up-down flow motion occurs along the right wall (see Fig. 6d and Fig. 10d). The flowfield for Case B4 has a strong down-up flow along the left wall, created by the motion of the bottom wall, while for Case C, the down-up flow is spread widely over the right half of the plane, and is relatively slow (compare Fig. 10d to Fig. 6d). The solid-body rotating central vortex is larger for Case B4 (compare Figs 10c,d to Figs 6c,d) that causes stronger convective mixing. For Case B4, big packets of unmixed fluid are formed at earlier time moments and reduced to relatively narrow areas of non-mixed fluid at T=20 (Fig. 10b).Therefore, the mixing rate is faster for Case B4 compared to Case C, and Case B4 is the second fastest to Case A in terms of mixing rate.

For mixing in the developed cavity flow, there is no slow diffusion-controlled phase of mixing for the first five time units as it were for developing flow (compare Fig 3b to Fig. 3a). After five time units, the initial interface between fluids has been ruptured and several areas with well-developed surface were formed (compare Fig. 11a to Fig. 5a). The mixing in all considered cases occurs in a faster pace compared to the developing flow (compare Fig. 3b to Fig. 3a). The completeness of mixing for developed flow after twenty units of time is larger than that for developing flow (compare Fig. 11b to Fig. 5b). The order of mixing rates for considered cases remains the same as for developing flow except for Case B4 that slightly outperforms Case A for T>30 (Fig. 3b).

## 4. Three-dimensional effects on mixing rate

In Figure 12, the flowfield and concentration of species are shown in the vertical plane perpendicular to the axis $x$ at $x=0.5$ in order to depict the three-dimensional effects in Cases C, B2, B3 and B4, including the finiteness of cavity in the z direction. The purpose is to show non-uniformity of mixing in the third spatial direction. The plane *(y,z)* at *x=0.5* is perpendicular to the x direction of lid motion and, therefore, shows the three-dimensional effects. In the absence of 3-D effects, concentration in this plane would be uniform in the z direction. The cases A[9] and B1[11] are genuinely there-dimensional and are not shown in Fig. 12 as their flowfield were discussed in Refs[9,11]. For example, for Case A the moving cavity bottom in the z direction creates strong flow as shown in Ref[9] to confirm that the flow is essentially three-dimensional.

For Case C the flow is nearly 2-D repeating itself in the z direction except for the areas near side walls. The area with high concentration of the first fluid is formed near the centerline at the lower part of cavity (*y<0.5*), while the area of high concentration of the second fluid is formed near the top of the cavity (Fig. 12a). Along the side walls (*z* close to *1* or *z* close to *0*), areas of higher concentration of the second fluid are formed at *y~0.5* and *y~0.25*, nevertheless, the fluids near the side walls are mixed to a larger degree (green regions in Fig. 12a). This shows the non-uniformity of concentration in the z direction.

For Case B2, the interface between two fluids is formed at *y~0.5*. The mixing forms local well-mixed regions of relatively small size near the interface (green areas in Fig. 12b). However, much larger areas of non-mixed first fluid (red area in Fig. 12b) and second fluid (blue area in Fig. 12b) are formed within top and bottom parts of cavity, respectively. The mixing rate for Case B2 is slower compared to other considered cases except Case B3 (Fig. 3a). The mixing toward the sides of cavity is closer to completion.

For Case B3 (Fig. 12c), the symmetry line is formed at *y=0.5* (see previous section). Local green areas of completed mixing are formed above and below the interface. The pairs of unmixed areas of an elliptic shape (red and blue) are formed in the interior of upper and lower halves of cavity. The formation of unmixed zones and slow mixing across the interface confirms the slowest mixing rate for Case B3 compared to the other considered cases (Fig 3a). Partially mixed regions are formed near the side walls of the cavity.

For Case B4 (Fig. 12d), the central partially-mixed (yellow) area surrounds unmixed (red) area in the lower half of the cavity. Symmetric mixing situation occurs for the upper half of the cavity. Relatively well-mixed areas are formed near the side walls (green and yellow areas) and at the bottom corners. To conclude this section, the presence of side walls (z close to 1 or z close to 0) does assist in mixing for cases C, B2, B3 and B4.

## 5. Mixing for large and small Schmidt numbers

The dynamics of deviation of concentration for pure convective mixing with zero diffusivity, $Sc \to \infty$, is shown in Fig. 3c. The mixing in Case A appears to be the fastest as it were for *Sc=1*. Case B2 is the second fastest to Case A for $Sc \to \infty$ because the process of changing the flowfield symmetry line from horizontal at T=0 to diagonal at T=20 is chiefly convective. The deviation of concentration for Case B3 is constant in time (no mixing) as the mixing in this case requires diffusion across the symmetry line, which does not exist for Sc→∞. For Case C, the mixing is the second slowest after B3 for Sc→∞.

The temporal rate of mixing, $(1/\sigma_0)\, d\sigma/dt$, is compared for cases $Sc=1$ and $Sc\to\infty$ in Fig. 4. For Case A, the maximum temporal rate of mixing is more than 50% larger for $Sc=1$ and occurs at $T\sim 10$ (Fig. 4a) compared to the maximum mixing rate at $T\sim 17$ for $Sc\to\infty$ (Fig. 4b). For other cases the maximum mixing rate is 50-100% larger for $Sc=1$ compared to Sc→∞. In the latter case the maximum occurs at a later time moment. For cases B4 and C the maximum mixing rate is comparable to that for Case A for $Sc=1$ while for $Sc\to\infty$ the maximum mixing rate is almost doubled for Case A compared to B4 and C. After five time units the interface surface is not ruptured for $Sc\to\infty$ (Fig. 13a). The minimum mixing rate in Fig. 4 is larger for Sc=1 compared to Sc→∞ (~1.25 versus ~0.25).

At T=20, the mixing for Case A (Sc→∞) lead to formation of multiple non-mixed zones with quite complex topology and well-developed interface (Fig. 13b). If physical diffusion of components is present, the well-developed interface surface would help to mix the fluids.

For the diffusion dominated mixing Sc→0, the thickness of diffusion mixing through the interface of mixing components can be evaluated by analytical solution for thickness δ of impulsively started diffusion boundary layer developing in the stationary fluid

$$\delta = \sqrt{\frac{T\mu}{\rho}} = \sqrt{\frac{T}{Re}} \quad (5)$$

For T=20, the thickness of the interface, $2\delta$, is equal to $2\sqrt{\frac{20}{2000}} = 0.2 \ll 1$. Therefore, the diffusion mixing is much slower/more local compared to convection mixing, which covers the entire cavity of the normalized size of unity.

For T=5, the thickness of the interface, $2\sqrt{\frac{5}{2000}} = 0.1$ that is confirmed by evaluation of approximate thickness of the interface, where the concentration increases from 0 to 1, in Fig. 5a. The interface thickness is much thinner for $Sc \to \infty$ compared to $Sc=1$ as diffusion is limited to numerical diffusion in the former case (compare Fig. 13a to Fig. 5a).

The characteristic convective time scale[32] can be chosen as $\delta/V$, where V is the velocity component perpendicular to interface between fluids. This is the time required for the fluid to be convected a distance of the order of thickness of the interface. The diffusive time scale is

$\delta^2/(D\ r)$, where $D = 1/(Sc\ Re)$ is the mass diffusivity and $r$ is the difference in concentration across the interface. Convective time scale increases with the time, T, as $\delta$ is proportional to $\sqrt{T}$ (see above). At T=5 (Fig. 5a) the interface is nearly normal to up-down developing velocity field in cavity while at later times the interface deviates from its normal orientation to local flow direction (Figs. 5b,d), that reduce V. The diffusive time scale increases with time as $\delta^2$ proportional to T and the difference in concentration across the interface decreases with time because of mixing (compare Fig. 5a to Fig. 5b). This explain the decay in mixing rate at time T approaches 20 (Figs. 3-4).

## 6. Mixing of fluids with different densities

Mixing of heavy and light fluids is considered in this section, where the density ratio is taken equal to two. Governing equations in this section are the three-dimensional compressible transient continuity and Navier-Stokes equations:

$$\frac{\partial \rho}{\partial t} + \frac{\partial(\rho u)}{\partial x} + \frac{\partial(\rho v)}{\partial y} + \frac{\partial(\rho w)}{\partial z} = 0,$$

(6)

$$\frac{\partial u}{\partial t} + u\frac{\partial u}{\partial x} + v\frac{\partial u}{\partial y} + w\frac{\partial u}{\partial z} + \frac{1}{\rho}\frac{\partial p}{\partial x} - \frac{1}{\rho Re}\left(\nabla^2 u + \frac{1}{3}\frac{\partial}{\partial x}(\nabla \cdot \vec{V})\right) = 0,$$

$$\frac{\partial v}{\partial t} + u\frac{\partial v}{\partial x} + v\frac{\partial v}{\partial y} + w\frac{\partial v}{\partial z} + \frac{1}{\rho}\frac{\partial p}{\partial y} - \frac{1}{\rho Re}\left(\nabla^2 v + \frac{1}{3}\frac{\partial}{\partial y}(\nabla \cdot \vec{V})\right) = 0,$$

$$\frac{\partial w}{\partial t} + u\frac{\partial w}{\partial x} + v\frac{\partial w}{\partial y} + w\frac{\partial w}{\partial z} + \frac{1}{\rho}\frac{\partial p}{\partial z} - \frac{1}{\rho Re}\left(\nabla^2 w + \frac{1}{3}\frac{\partial}{\partial z}(\nabla \cdot \vec{V})\right) = 0,$$

(7)

$$\frac{\partial c}{\partial t} + u\frac{\partial c}{\partial x} + v\frac{\partial c}{\partial y} + w\frac{\partial c}{\partial z} - \frac{1}{\rho Sc\ Re}\nabla^2 c = 0,$$

(8)

where $\vec{V}=(u,v,w)$ is the velocity vector field.

For the cases with density ratio 2:1 the upper half of cavity is initially filled with the fluid having doubled density compared to that of the lower half. For the version of case C with density ratio 1:2 the upper half of cavity (y>0.5) is initially filled with the lighter fluid. To have the same mass of fluid in cavity as in prior sections, in which density was taken equal to unity, the density ρ of the lighter fluid is selected by equation (ρ+2ρ)/2=1 so as the densities of the light and heavy fluids are 2/3 and 4/3, respectively. The viscosity, μ, and diffusivity, D, of both fluids remains equal. The kinematic viscosity becomes different for two fluids because of their different densities and, consequently, the Schmidt numbers for light and heavy fluids will be equal to $\frac{3}{2}$ and $\frac{3}{4}$, respectively. The variance of concentration, Eq. (4), is modified as for non-equal densities for completed mixing the concentration (mass fraction) of the first fluid should be equal to $\frac{\rho_1}{\rho_1+\rho_2}$:

$$\sigma^2 = (\int \left(c - \frac{\rho_1}{\rho_1+\rho_2}\right)^2 dV)/V \ , (5)$$

where $\rho_1$ is the fluid initially filling the upper half of the cavity.

Initial value of $\sigma^2$ is equal to $\sigma_0^2$, where $\sigma_0^2 = 1/2\left(\left(1 - \frac{\rho_1}{\rho_1+\rho_2}\right)^2 + \left(0 - \frac{\rho_1}{\rho_1+\rho_2}\right)^2\right)$. The value of normalized deviation of concentration, $\sigma/\sigma_0$, as a function of time is shown in Fig. 14.

The dynamics of normalized deviation of concentration for the prior cases A and C (Fig. 3a) are presented in Fig. 14a for comparison. The mixing for fluids with different densities appear to be faster compared to corresponding cases A and C in which the density was equal to unity. While for case A the mixing rate does not depend on whether the heavy fluid initially occupies lower or upper half of cavity, for case C the cavity flow is single (upper) lid driven and, therefore, the mixing rate is different for these two situations. Nevertheless, the difference in mixing rate between these two situations is much smaller compared to the difference between either of the situations and the case C with uniform density of fluid (Fig. 14a).

To confirm that the density ratio is the lead reason for elevated mixing rate, the Case A was modeled with uniform density and $Sc=3/4$ (see Fig. 14a). As soon as $Sc<1$, the diffusivity becomes larger than kinematic viscosity and, therefore, the mixing is faster compared to the original Case A. The difference in mixing rate between original Case A (with $\rho=1$ and $Sc=1$) and the modified Case A with $Sc=0.75$ is minor compared to the difference between original Case A and Case A with density ratio 2:1.

At T=20, the rates of mixing from fastest to slowest are ranked A, B1, C, B2, B4, B3, where the difference in mixing rate between C, B2 and B4 is small. The order of rates of mixing remains similar to that for cases with uniform density except for case B1 that outperforms cases C and B4. The genuinely 3-D cases A and B1 have the fastest mixing for non-equal density fluids.

In Fig. 14b, the maximum mixing rate, $d(\sigma/\sigma_o)/dt$, reaches the value of 16 for the fastest mixing Cases A and B1. The magnitude of maximum of mixing rate is more than doubled compared to corresponding cavity flow case with uniform density (Fig. 4a).

In Fig. 15a the interface between fluids is shown in Case A with density ratio 2:1 at T=5. Compared to the Case A with uniform density (Fig. 5a), the deformation of interface is more pronounced and the interface is shifted farther from its initial location at y=0.5. The divergence of the velocity field, $\nabla \cdot \vec{V}$, which is the rate of volume expansion affecting velocity vector field by Eq. (7), is shown in Fig. 15b. This term is equal to zero when density is constant. The figure shows substantial magnitude of this term near the interface of two fluids (see Figs. 15a and b). The magnitude of $\nabla \cdot \vec{V}$ is larger at the right part of cavity where the velocity is larger (see Fig. 15c), the interface zone is thinner (see Fig. 15a) and, consequently, the density and concentration gradient is larger compared to the left part of cavity. The presence of non-zero divergence term in the right hand side of Eq. (7) increase the velocity at fluids' interface compared to Case A with uniform density (compare the isolines of velocity magnitude in Figs. 15c,d) and thus enhance mixing of two fluids.

The distribution of vorticity for mixing of heavy and light fluids is depicted in Fig. 16 for cases A and C. For Case A the zones of large vorticity are formed in the peripheral part of rigidly rotating vortex (compare Fig. 5c and Fig. 16 a). For Case C, the similar vorticity spots are formed for mixing of fluids with non-equal densities (compare Fig. 16b to Fig 6c) although the strength of these vorticity spots is not as prominent as for Case A (compare Fig. 16a to Fig. 16b).

**Conclusions**

The mixing rate is obtained numerically for 3-D wall-driven cavity flows. The dynamics of mixing rate allow to determine two phases of mixing: (i) slow mixing caused by diffusion through deforming interface during first five time units; and (ii) fast mixing rate caused by convection and development of multiple interfaces. The maximum of mixing rate is reached within the time interval $5 \leq T \leq 20$.

The six different modeled set-ups include the cavity with its top and bottom walls moving in perpendicular directions (Case A), the cavity with its lid moving within its plane along its diagonal (Case B1), the cavity with its top and one of its side vertical walls moving in perpendicular directions (Case B2), the cavity in which the top and bottom walls move parallel to each other (Case B3), and the cavity in which its top and bottom walls move in opposite directions (Case B4). The mixing of two fluids of equal density and viscosity initially occupying the upper and lower halves of cavity is modeled so that the variance of concentration is computed as a function of time.

The mixing rates for the cases listed above are compared to the baseline single lid driven cavity flow (Case C).

The intensity of mixing was evaluated first for developing cavity flow, which was initially at rest, for twenty time units and developed flow for next twenty time units. All set-ups have quite similar slow mixing rates for the first five time units except for the faster mixing case B2. During the first time units the separating interface between two fluids is deformed but not ruptured yet and mixing is by diffusion through the interface. For twenty units of mixing time the mixing rate is the fastest for Case A followed by Case B4, moderate mixing rate cases B1, B2 and C, and slow-mixing B3. For the developed cavity flow, there is no slow mixing phase during the first five units. Similar to the developing flow, the faster-mixing cases are B4 and A.

The interface between two fluids for the Case A is significantly more developed compared to the interface between two fluids for Case C. The genuinely 3-D flow in Case A leads to the formation of many smaller size areas of non-mixed fluids, which causes faster mixing. For the limit case of pure convection mixing, $Sc \to \infty$, formation of complex topology of mixing fluids with well-developed multiple interfaces is shown. Case B4 has a relatively fast mixing rate compared to other cases because of stronger rotational motion of central vortex in the *(x, y)* plane. The 3-D effects in the *(z, y)* plane assist to speed-up mixing for cases with essentially 2-D flowfield (B2, B3, B4 and C).

Case B1 and Case C have comparable mixing rate at later stages of mixing. For Case B1, mixing at the central diagonal rectangular cavity, which is relatively shallow $\sqrt{2}:1$, is more intensive compared to the squared *(x, y)* cavity for Case C. However, for Case B1 mixing in off-diagonal cavities with larger depth-to-length ratios is relatively slow. The 3-D flow in the direction normal to the diagonal lid motion softens the above difference in mixing making it comparable to the baseline Case C.

Case B2 has the fastest mixing rate after beginning of motion of the top and left cavity walls. As time progresses, symmetrical flowfield is formed with respect to the cavity diagonal plane passing through the upper left corner. The transition of the interface between mixing fluids from the horizontal (at $y=0.5$) to diagonal plane of symmetry creates a fast mixing rate for two fluids. The mixing rate substantially slows down as soon as diagonal symmetry has formed.

Mixing of fluids with density ratio equal to two is considered. The mixing appears to be faster for all cases compared to the uniform density mixing. Mixing rate for the Case A clearly outperform other cases (of which Case B1 is the fastest) that confirm the usefulness of the set-up with genuinely 3-D mixing.

Dynamics of characteristic diffusion and convection times explains the slow pace of mixing when time approaches 20 units by analytical evaluation of interface thickness. The rate of volume expansion for mixing of fluids with different densities affects velocity vector field and leads to increase of penetration of velocity deeper into cavity that enhances mixing. The magnitude of maximum of mixing rate is more than doubled compared to corresponding cavity flow with uniform density. The difference in density of two mixing fluids appears to have larger effect on the rate of mixing compared to the difference in kinematic viscosity.

The presented evaluation of flow mixing intensity and rates for the various arrangements of moving cavity walls will assist in the evaluation of mixing in enclosures with moving and stationary parts.

## Appendix

## Validation of Numerical Methodology

The computational software ANSYS/Fluent was validated by comparison with prior numerical solutions for 2-D and 3-D cavity flows with a single moving wall. The 2-D cavity flow ($0<x<1$, $0<y<1$) has moving boundary ($y=1$, $0<x<1$) in the positive x direction with $U=1$ and $Re=1000$. The cavity domain is discretized using the 201 x 201 finite-volume grid with uniform grid step $h=0.05$. The finite-volume discretization of convective terms is second-order upwind. The steady state solution is obtained starting with initial condition $u=0.5$ and $v=0.02$ using SIMPLEC[22,24] method available in frame of ANSYS/Fluent. Relaxation parameters are 0.3 for continuity equation and 0.7 for momentum equations. Stopping criteria for residuals[22] are selected equal to $10^{-06}$. The numerical values of variables obtained in finite-volume cells are interpolated by Fluent to the points of comparison established in literature (see coordinates x and y along centerlines in Tables below).

Comparisons with other numerical methods for 2-D cavity are shown in Tables A-1 and A-2. Comparative studies used the vorticity-stream function formulation of the two-dimensional incompressible Navier-Stokes equations with multigrid finite-difference methods[33], spectral methods[34] discontinuous Galerkin finite-element method (DG-FEM)[35] and mesh-free Smoothed Particle Hydrodynamics (SPH)[36]. In some of the above listed references the coordinates' origin is located at cavity center, (0.5, 0.5), and coordinates are directed in the opposite way. In such cases their results are presented in terms of the current set-up of coordinate system and moving wall.

Comparison of velocity u, parallel to moving lid, along the centerline, $0 \leq y \leq 1$, to Refs[33,34] is shown in Table A-1. The current results are close to those published in Ref[34]. Note that Ref[36] does not present centerline velocity in tabular format but compared to Ref[33] graphically (see Fig. 9[36]) to show minor discrepancy between their results; Ref[35] shows graphical correspondence of computed centerline velocity to Refs[33,34]. Location and strength of primary vortex in 2-D cavity is shown in Table A-2. Presented computations compare most closely to results presented in Refs[33,35]. Ref[36] presents the range of results with different approximations of boundary conditions at cavity walls as reflected in Table A-2.

For the three-dimensional cavity flow (Case C in Fig.1) the velocity at the cavity's vertical and horizontal centerlines corresponds well to Ref[18] (see Tables A-3, A-4 and Fig. A1).

**Table A-1: Velocity u along vertical centerline of 2-D cavity**

| y-coordinate | Ghia[33], Table 1, p. 398, 1982 | Botella[34], Table 9, p. 426, 1998 | Current Fluent computations |
|---|---|---|---|
| 1.0000 | 1.0000 | 1.0000 | 1.0000 |
| 0.9766 | 0.65928 | 0.6644227 | 0.6641546 |
| 0.9688 | 0.57492 | 0.5808359 | 0.5804891 |
| 0.9609 | 0.51117 | 0.5169277 | 0.5166981 |
| 0.9531 | 0.46604 | 0.4723329 | 0.4722601 |
| 0.8516 | 0.33304 | 0.3372212 | 0.3364770 |
| 0.7344 | 0.18719 | 0.1886747 | 0.1888589 |
| 0.6172 | 0.05702 | 0.0570178 | 0.0568349 |
| 0.5000 | -0.06080 | -0.0620561 | -0.0621449 |
| 0.4531 | -0.10648 | -0.1081999 | -0.1082493 |
| 0.2813 | -0.27805 | -0.2803696 | -0.2803804 |
| 0.1719 | -0.38289 | -0.3885691 | -0.3861539 |
| 0.1016 | -0.29730 | -0.3004561 | -0.2984093 |
| 0.0703 | -0.22220 | -0.2228955 | -0.2209065 |
| 0.0625 | -0.20196 | -0.2023300 | -0.2003669 |
| 0.0547 | -0.18109 | -0.1812881 | -0.1794080 |
| 0.0000 | 0 | 0 | 0.0 |

**Table A-2: Location and strength of primary vortex in 2-D cavity**

| Primary vortex | Ghia[33], p.408, 1982 | Botella[34], Table 6, p. 425, 1998 | Khorasanizade[36], Table 4, p.665, 2014 | Romano[35], p. 493, 2017 | Current Fluent computations |
|---|---|---|---|---|---|
| x coordinate | 0.5313 | 0.5308 | 0.5307 to 0.5325 | 0.531 | 0.5328 |
| y coordinate | 0.5625 | 0.5652 | 0.5646 to 0.5676 | 0.564 | 0.5641 |
| streamfunction, $\|\psi\|$ | 0.1179 | 0.1189 | 0.0947 to 0.1143 | 0.118 | 0.1180 |
| vorticity, $\|\omega\|$ | 2.0497 | 2.0678 | 1.7185 to 2.0569 | 2.045 | 2.0424 |

**Table A-3: Velocity u along vertical centerline, (0.5, 0≤y≤1; 0.5) for 3-D cavity**

| Coordinate y[18] | Value of velocity u Albensoeder[18], Table 5, p. 548, 2005 | Computed value of velocity u by Fluent |
|---|---|---|
| 1 (at moving lid) | 1.0 | 1.0 |
| 0.9766 | 0.589641 | 0.597402 |
| 0.9688 | 0.484428 | 0.462556 |
| 0.9609 | 0.398209 | 0.407877 |
| 0.9531 | 0.331711 | 0.321799 |
| 0.8516 | 0.121829 | 0.124317 |
| 0.7344 | 0.0733444 | 0.073147 |
| 0.6172 | 0.0390483 | 0.039349 |
| 0.5 | 0.0080177 | 0.0072672 |
| 0.4531 | -0.0061192 | -0.0065531 |
| 0.2813 | -0.109989 | -0.111251 |
| 0.1719 | -0.251601 | -0.248257 |
| 0.1016 | -0.272929 | -0.272853 |
| 0.0703 | -0.236955 | -0.232010 |
| 0.0625 | -0.222826 | -0.222721 |
| 0.0547 | -0.206233 | -0.201010 |
| 0 (at stationary wall) | 0 | 0 |

**Table A-4: Velocity v along horizontal centerline, (0≤x≤1; 0.5; 0.5) ) for 3-D cavity**

| Coordinate $x^{18}$ | Value of velocity v Albensoeder[17], Table 6, p. 549, 2005 | Computed value of velocity v by Fluent |
|---|---|---|
| 1 (at front wall) | 0 | 0 |
| 0.9688 | -0.188642 | -0.194341 |
| 0.9609 | -0.240947 | -0.227352 |
| 0.9531 | -0.290317 | -0.290717 |
| 0.9453 | -0.335112 | -0.319831 |
| 0.9063 | -0.434231 | -0.431648 |
| 0.8594 | -0.311172 | -0.319249 |
| 0.8047 | -0.152230 | -0.155792 |
| 0.5 | 0.0367355 | 0.0376096 |
| 0.2344 | 0.169866 | 0.169041 |
| 0.2266 | 0.175803 | 0.176693 |
| 0.1563 | 0.229236 | 0.229667 |
| 0.0938 | 0.244074 | 0.242652 |
| 0.0781 | 0.235027 | 0.233922 |
| 0.0703 | 0.227462 | 0.223813 |
| 0.0625 | 0.217384 | 0.217183 |
| 0 (at back wall) | 0 | 0 |

# Figures

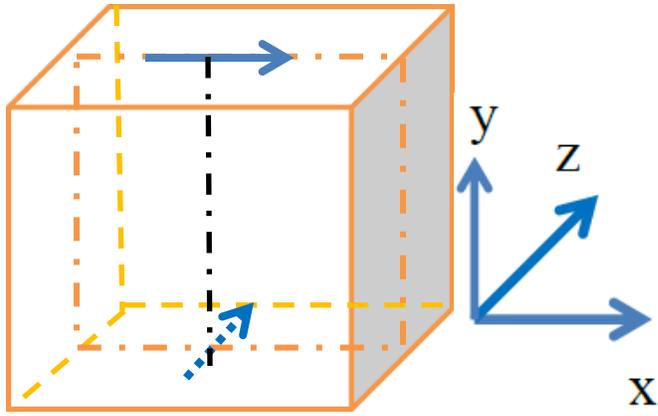
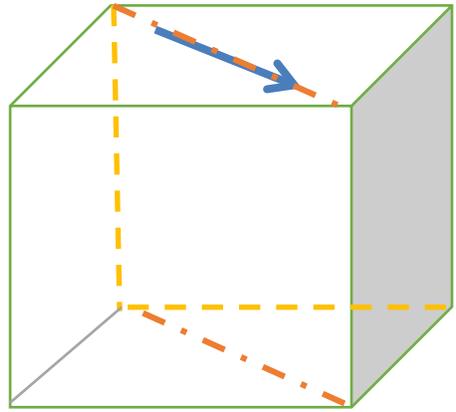

(a)

(b)

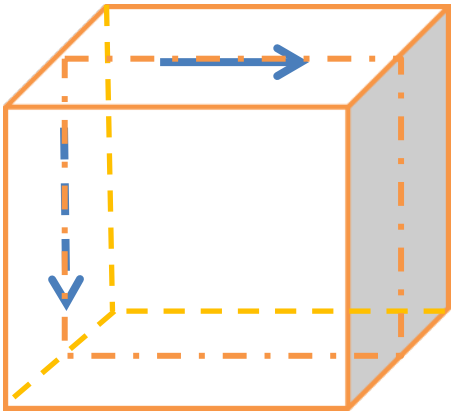
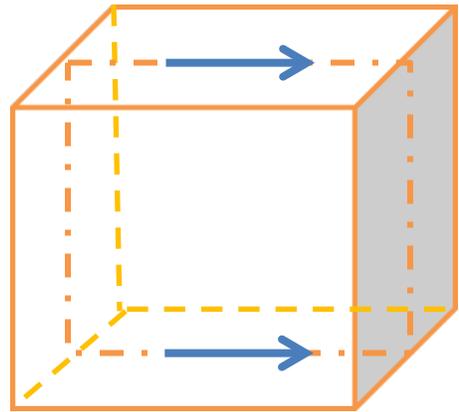

(c)

(d)

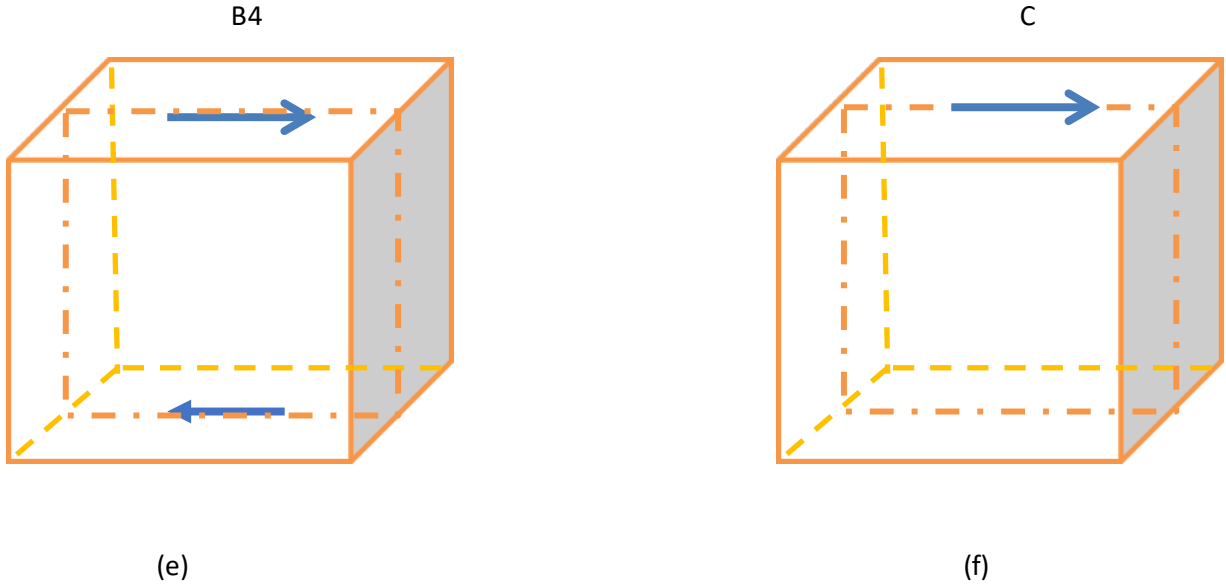

(e)  (f)

Figure 1: Flows induced by moving walls of cubical cavity: (a) top and bottom walls move in perpendicular directions (Case A), (b) top wall moves along its diagonal (Case B1), (c) top wall moves along the x direction and the left wall moves down in the negative y direction (Case B2), (d) top and bottom walls move parallel (Case B3), (e) top and bottom walls are in reverse motion (Case B4) and (f) top wall moves along its edge (Case C). The central vertical plane is shown in which flowfield and concentration are depicted.

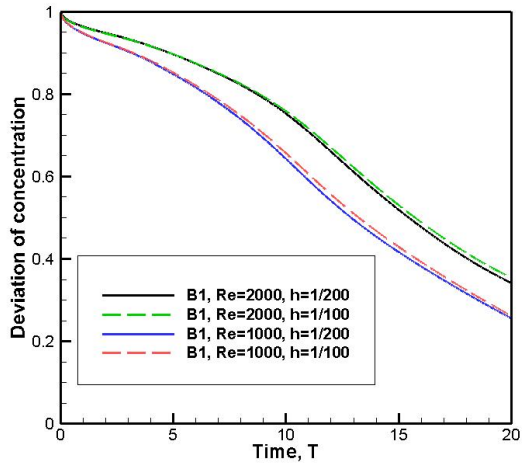
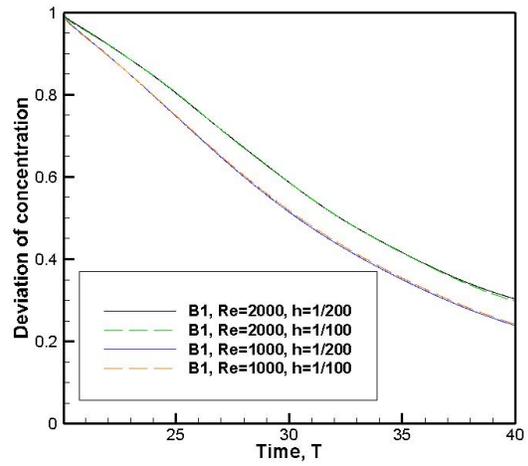

(a)            (b)

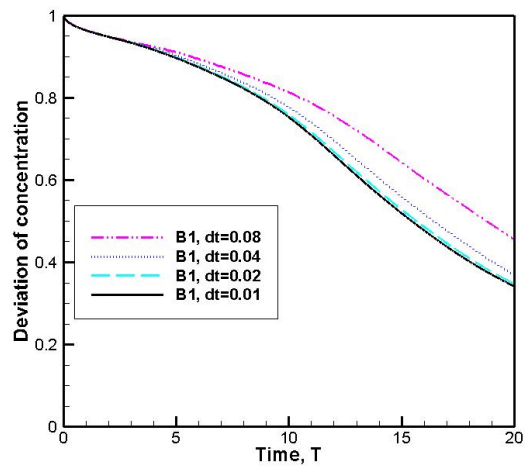

(c)

**Figure 2:** Grid convergence for Case B-1: a) developing flow (0≤T≤20), b) developed flow (20≤T≤40) and c) effect of time step, Δt, on developing flow for $201^3$ gird. Finite-volume grids $201^2$ and $101^2$ are used in (a) and (b) for comparison.

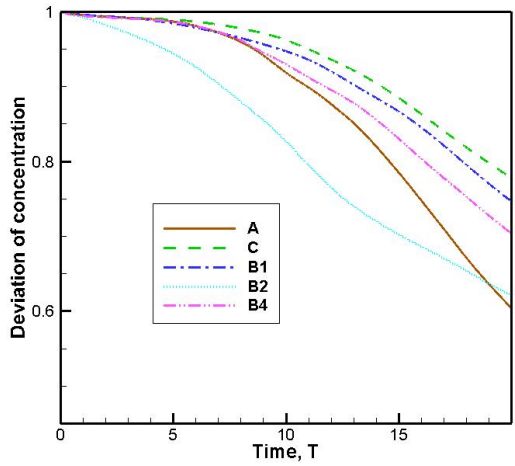
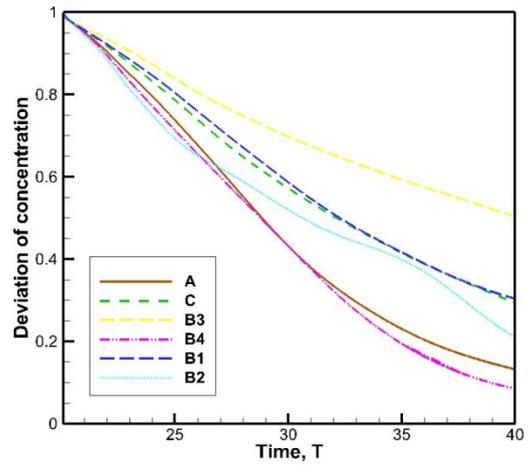

(a) (b)

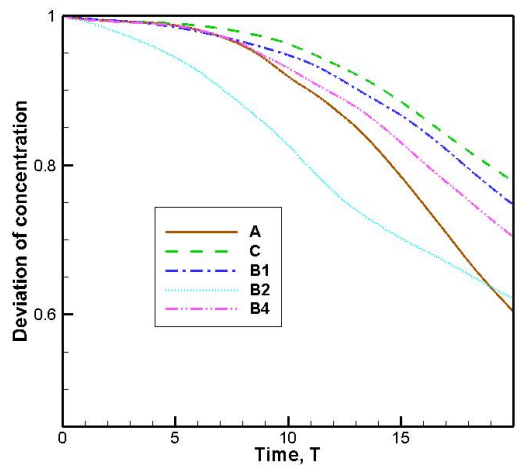

(c)

Figure 3: Deviation of concentration, σ, for cases A, B1-B4 and C: (a) developing cavity flow, (b) developed cavity flow with the concentration re-set at T=20 and (c) pure convective mixing in developing flow (no diffusivity, Sc→∞). For (c) there is no mixing in Case B3.

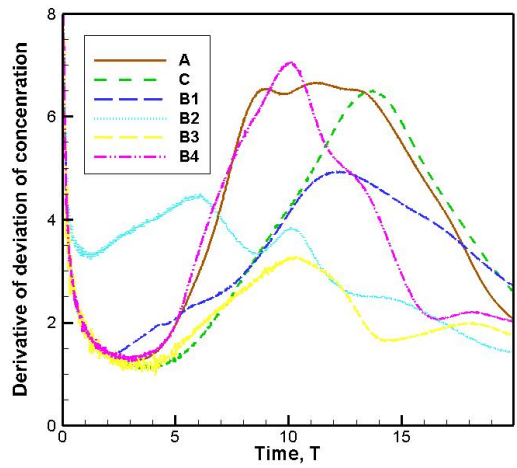 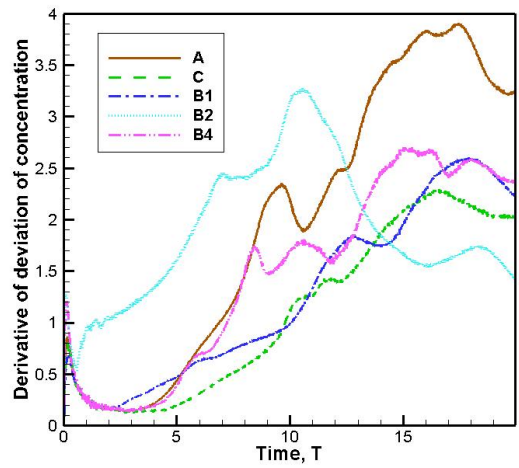

(a)             (b)

Figure 4: Time rate of deviation of concentration, $d(\sigma/\sigma_o)/dt$: a) Sc=1 and b) no diffusivity, Sc→∞

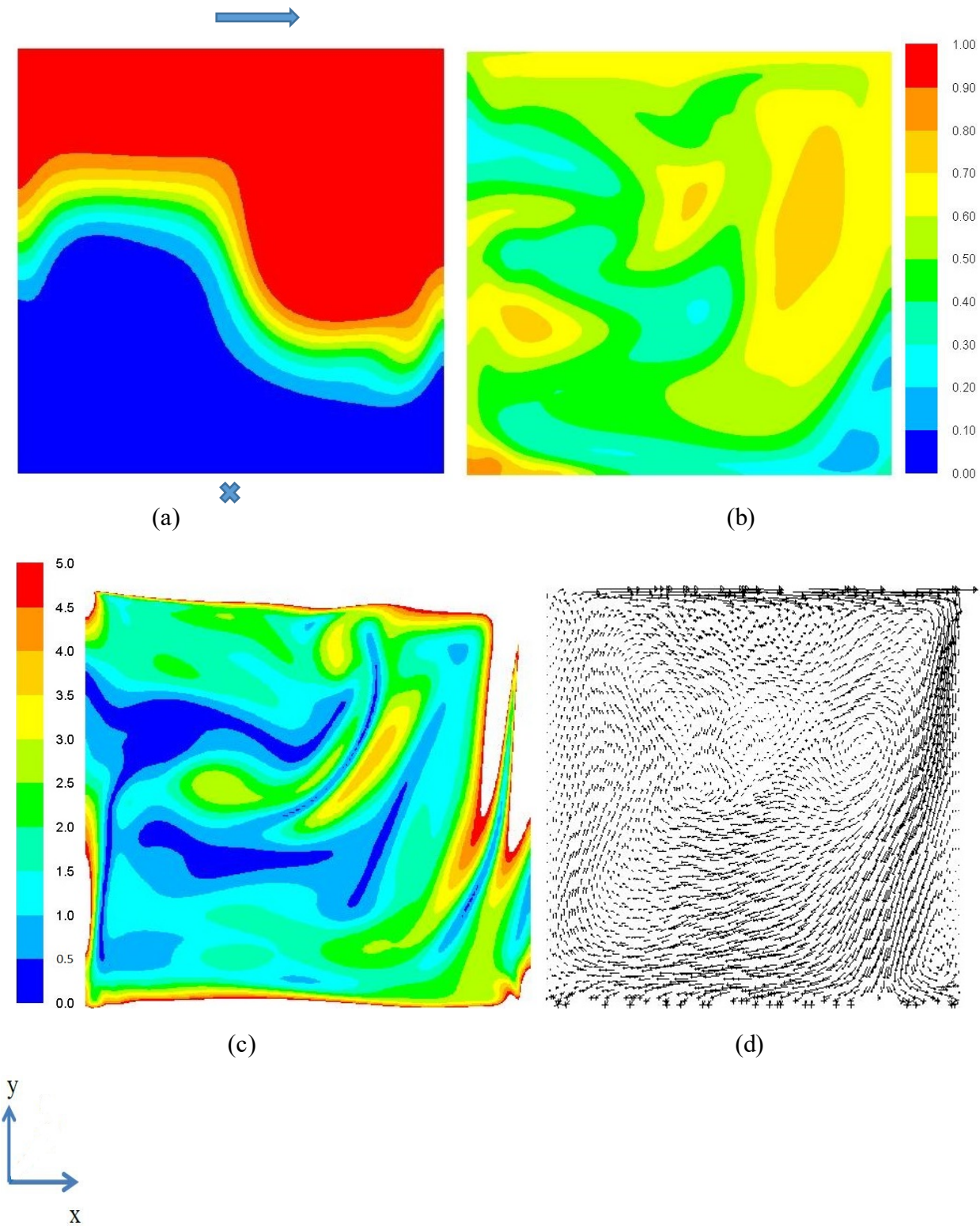

Figure 5: Case A (developing flow): concentration (mass fraction) of the first fluid at time moments (a) 5 ,(b) 20 units, (c) vorticity magnitude and (d) velocity vector field at 20 time units. In the current figure and in all following figures, concentration is plot with 10 color sub-

divisions, where the concentration maximum is equal to 1 and minimum is 0. Vorticity magnitude is plot using 10 color sub-divisions for the vorticity magnitude interval $0 \leq |\omega| \leq 5$.

✖    is the moving wall direction, which is perpendicular to the figure plane

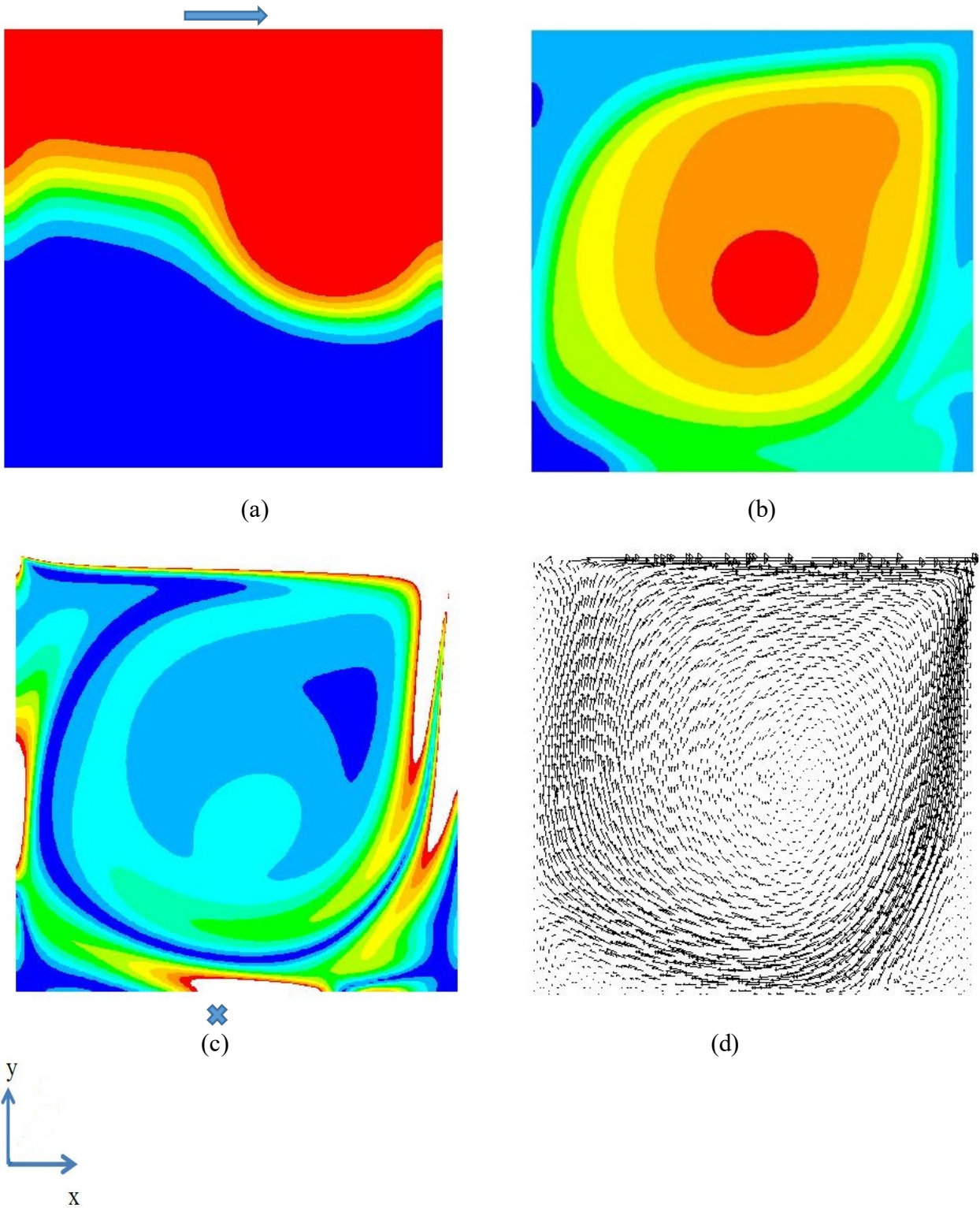

Figure 6: Case C: conditions are the same as in Fig. 5

✖ is the moving wall direction, which is perpendicular to figure plane

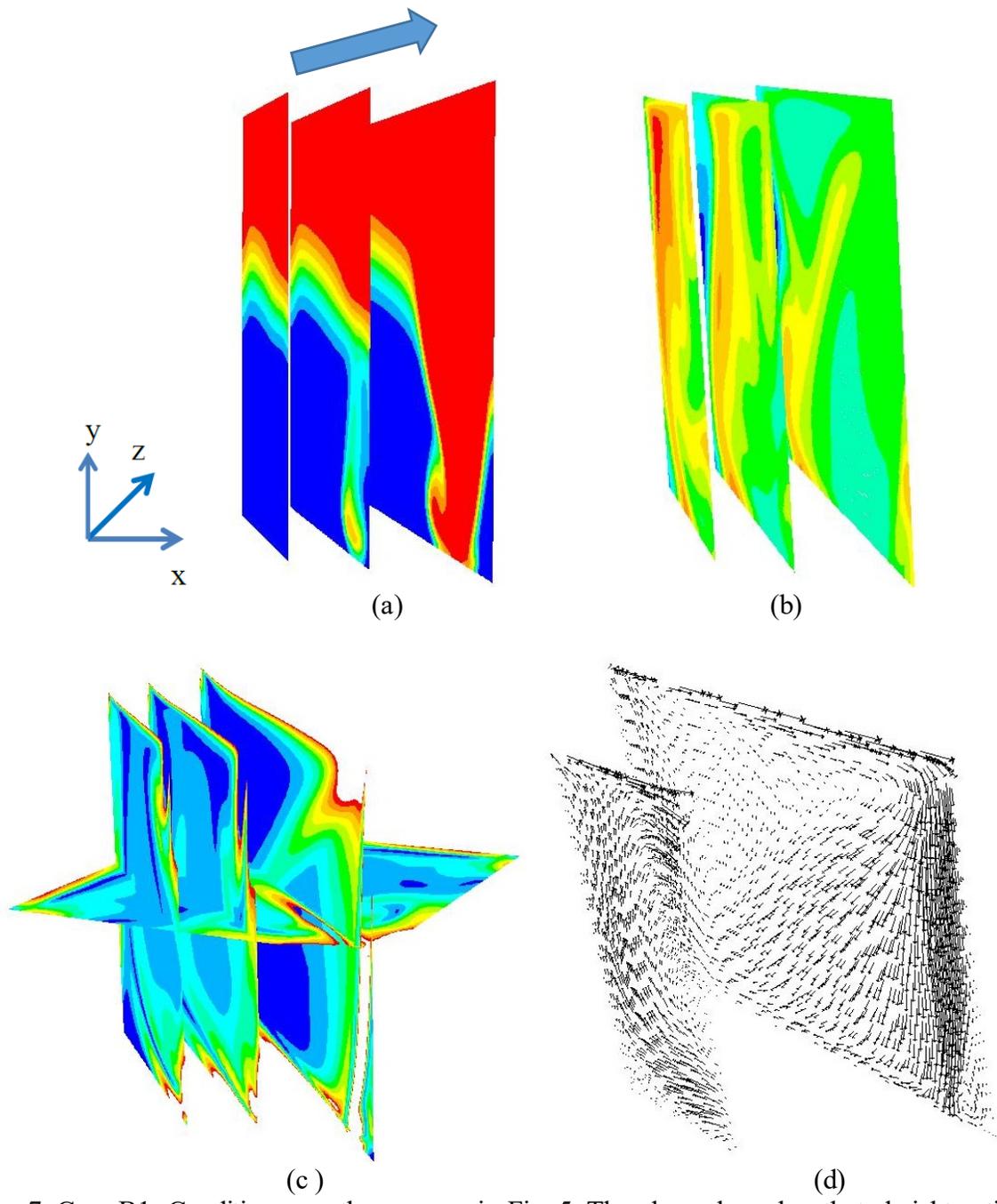

Figure 7: Case B1: Conditions are the same as in Fig. 5. The planes have length to height ratio √2 : 1, 1: 1 and ½:1. The plane 1:1 is not shown in (d)

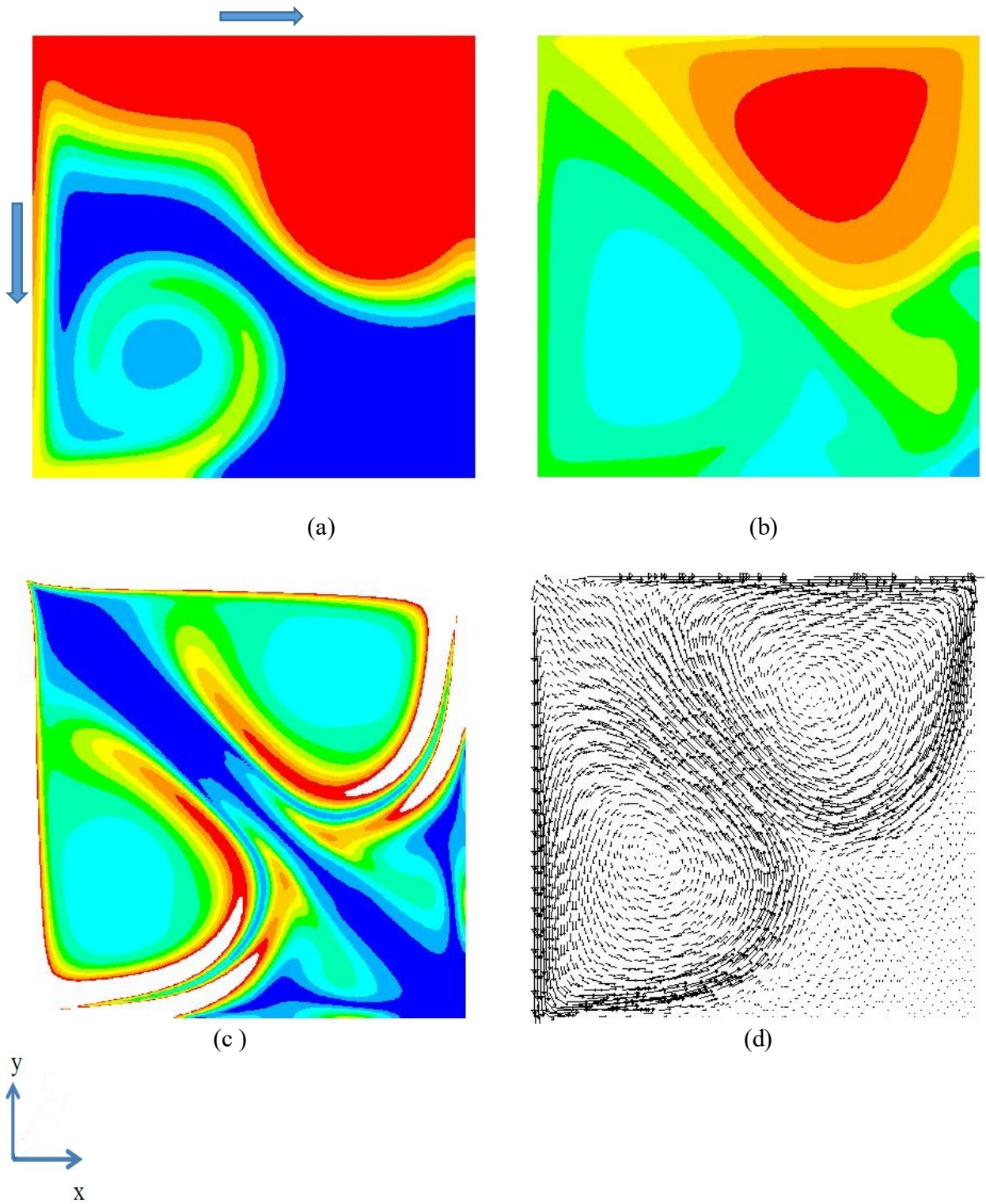

Figure 8: Case B2: conditions are the same as in Fig. 5

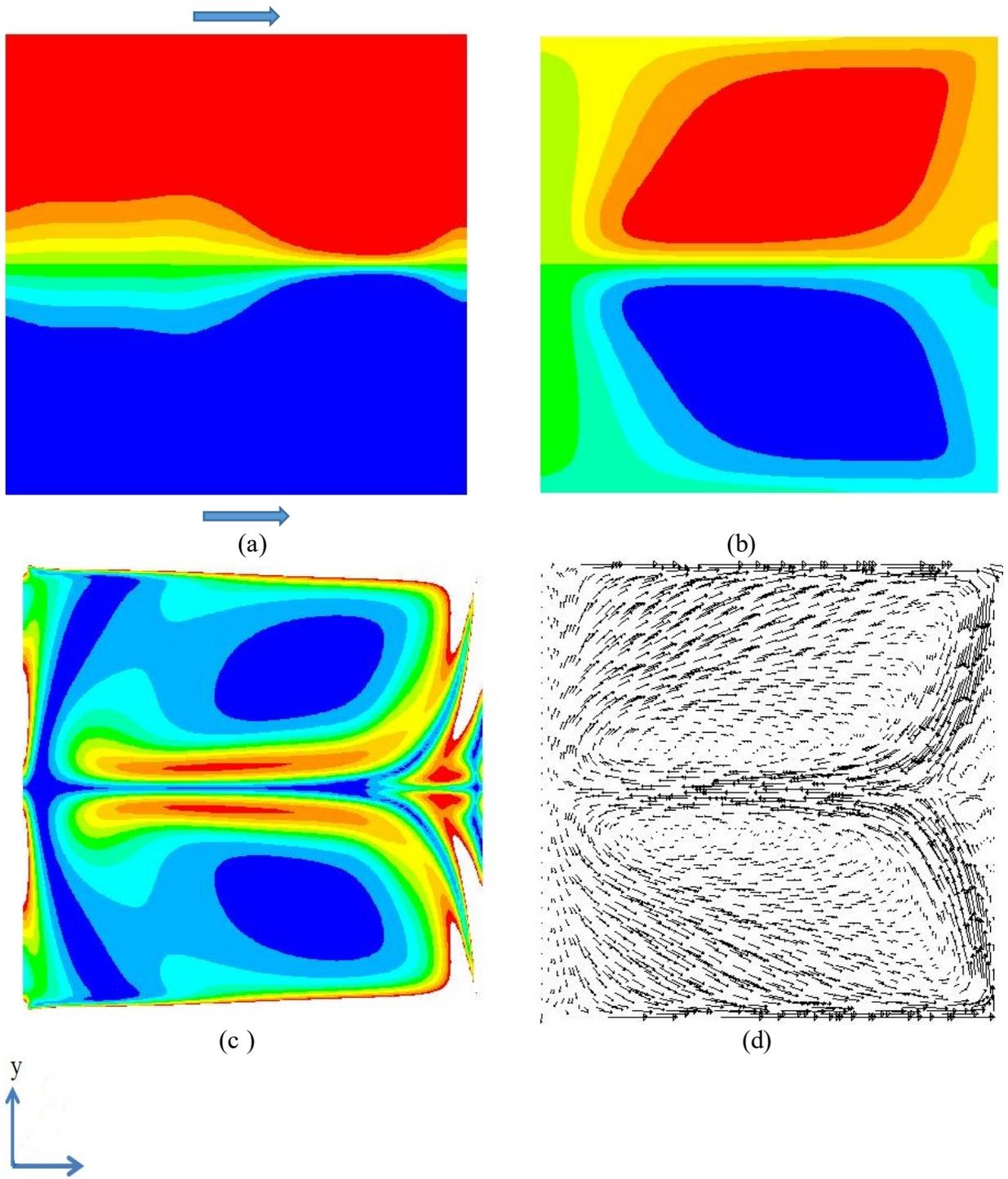

Figure 9: Case B3. Conditions are the same as in Fig. 5.

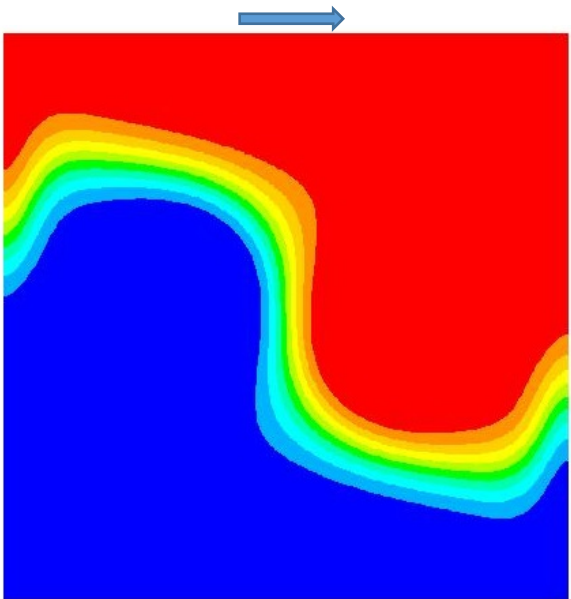
(a)

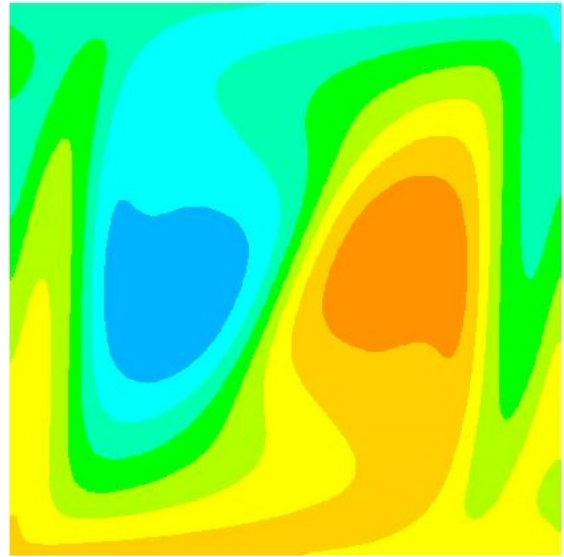
(b)

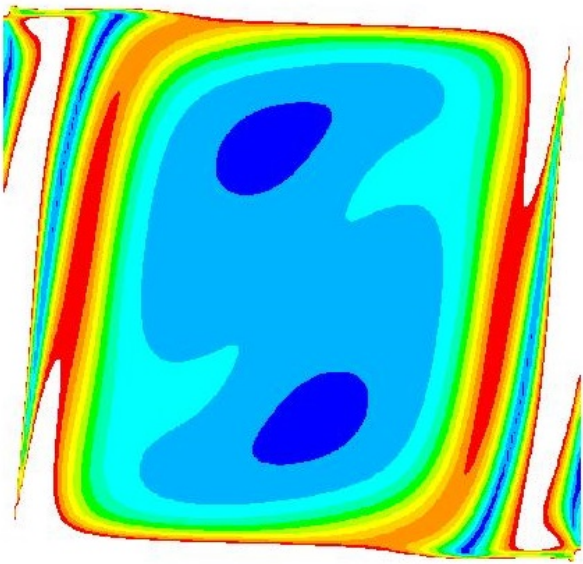
(c)

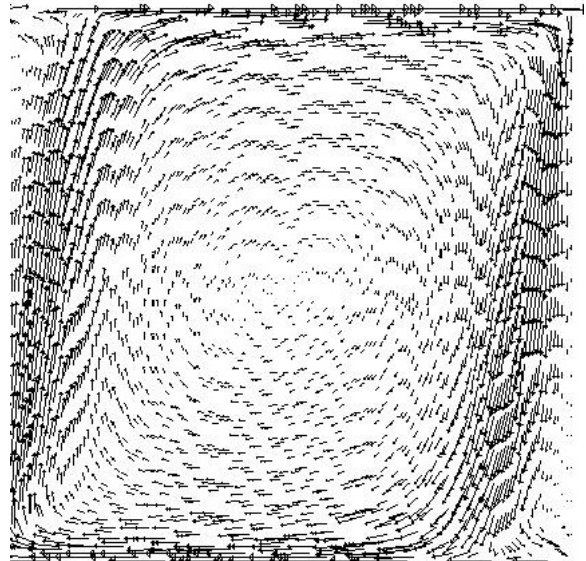
(d)

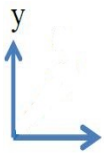

Figure 10: Case B4. Conditions are the same as in Fig. 5.

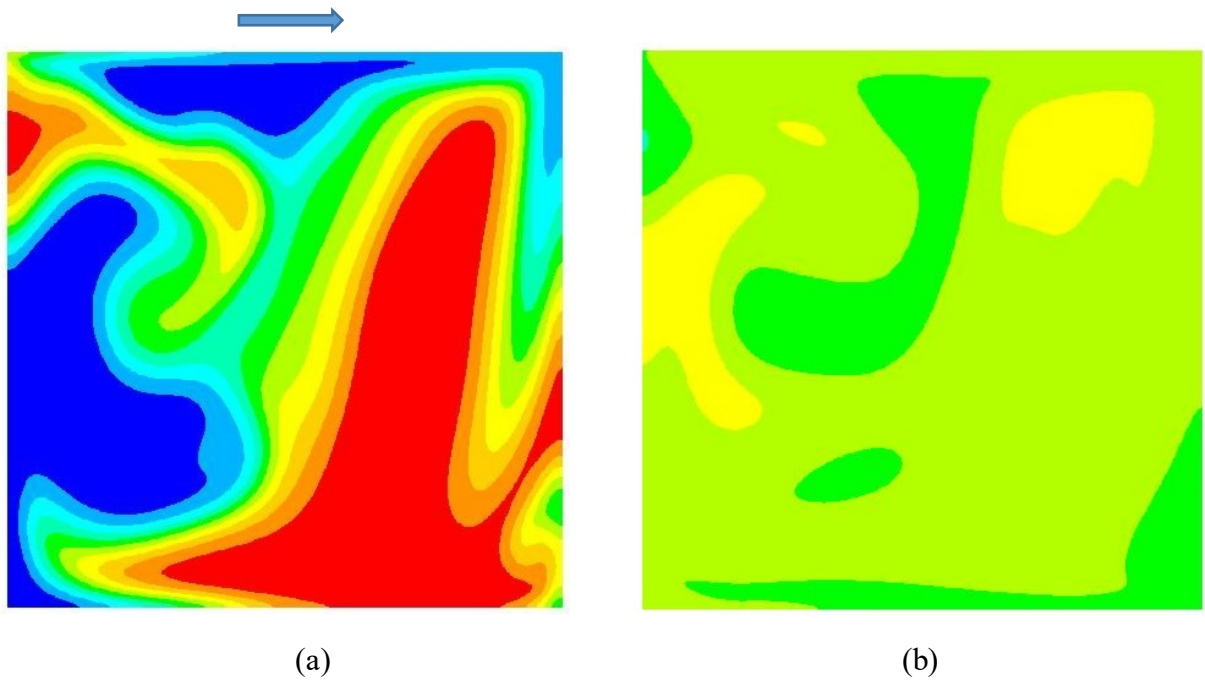

(a)                                    (b)

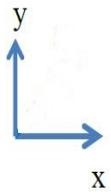

Figure 11: Case A (developed flow): concentration of first fluid at time moments (a) 5 and (b) 20 units after concentration re-set.

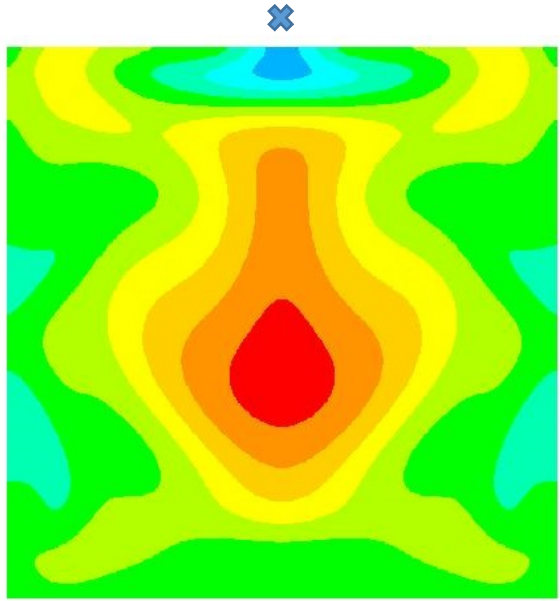

(a)

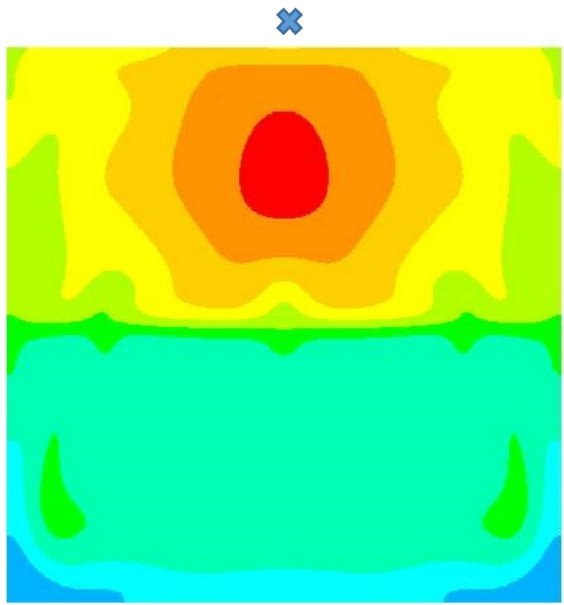

(b)

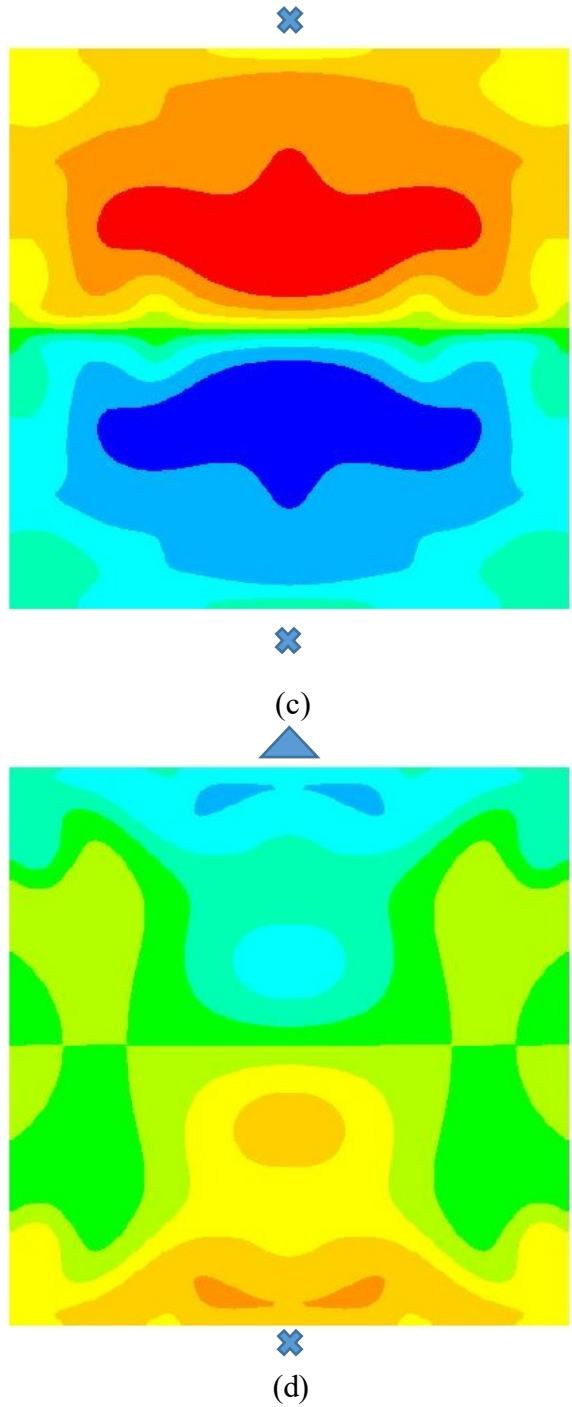

Figure 12: Concentration and flowfield in vertical (y,z) plane, *x=0.5* at *T=20*: (a) Case C, (b) Case B2, (c) Case B3, and (d) Case B4.

✖   the moving wall direction perpendicular to the figure plane directed toward the back

▲   the moving wall direction directed toward the front.

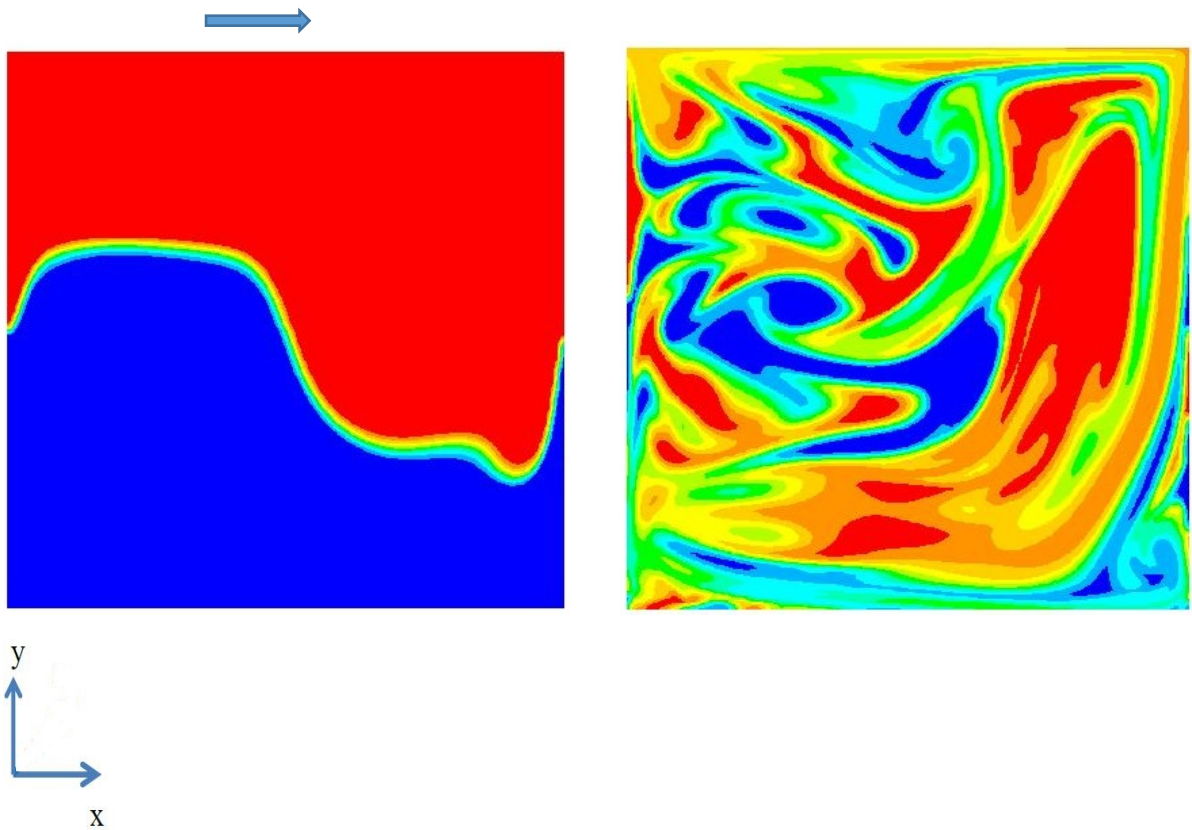

Figure 13: Case A: Concentration of first fluid for zero diffusion (Sc→∞) mixing in developing cavity flow: (a) 5 and (b) 20 units.

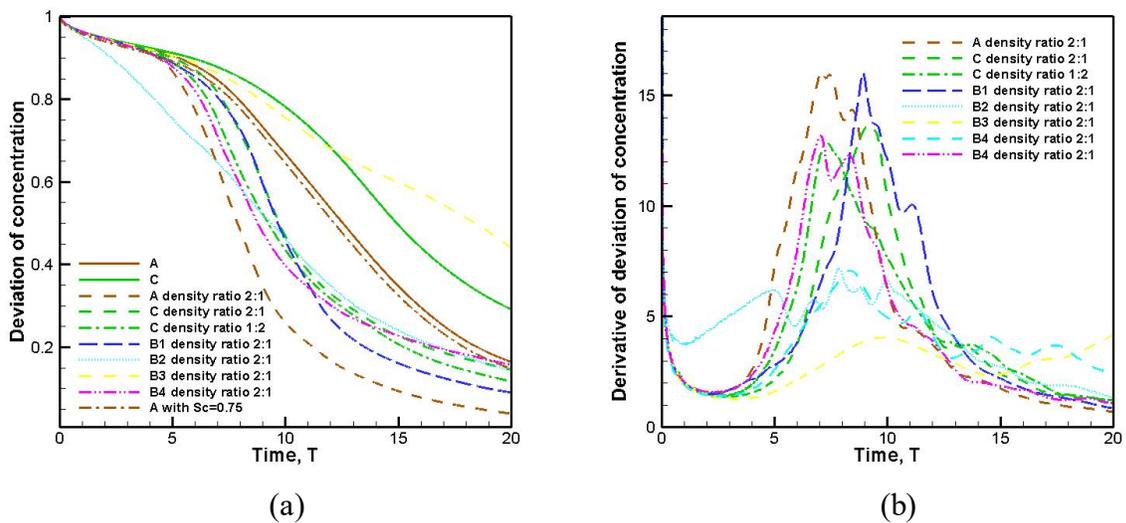

Figure 14: mixing of heavy and light fluids (a) deviation of concentration and (b) temporal derivative of deviation of concentration

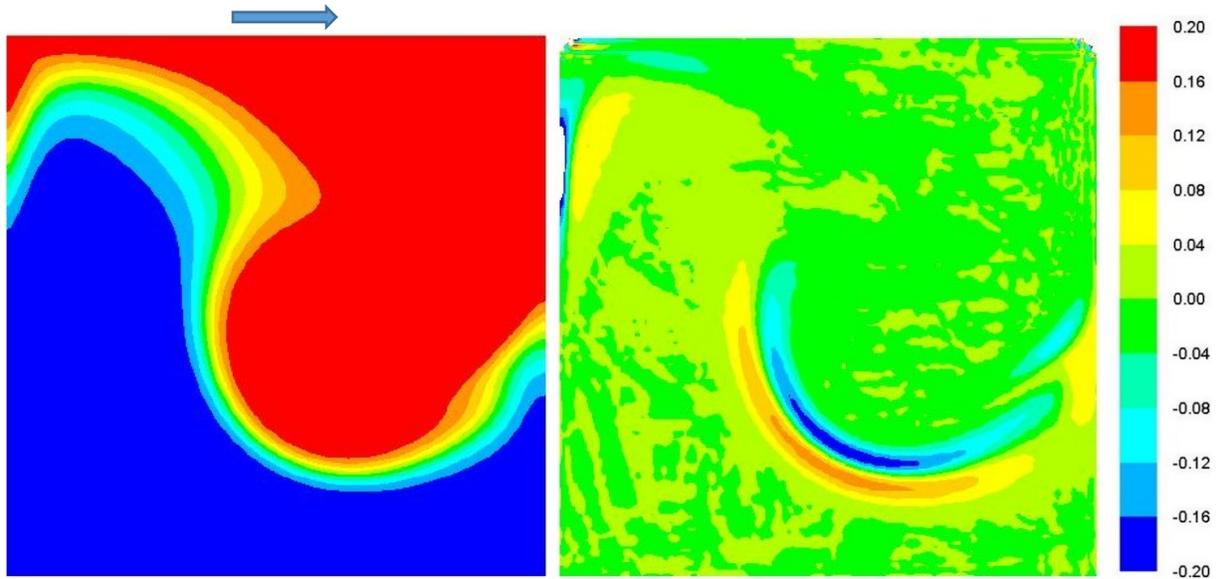

(a)                  (b)

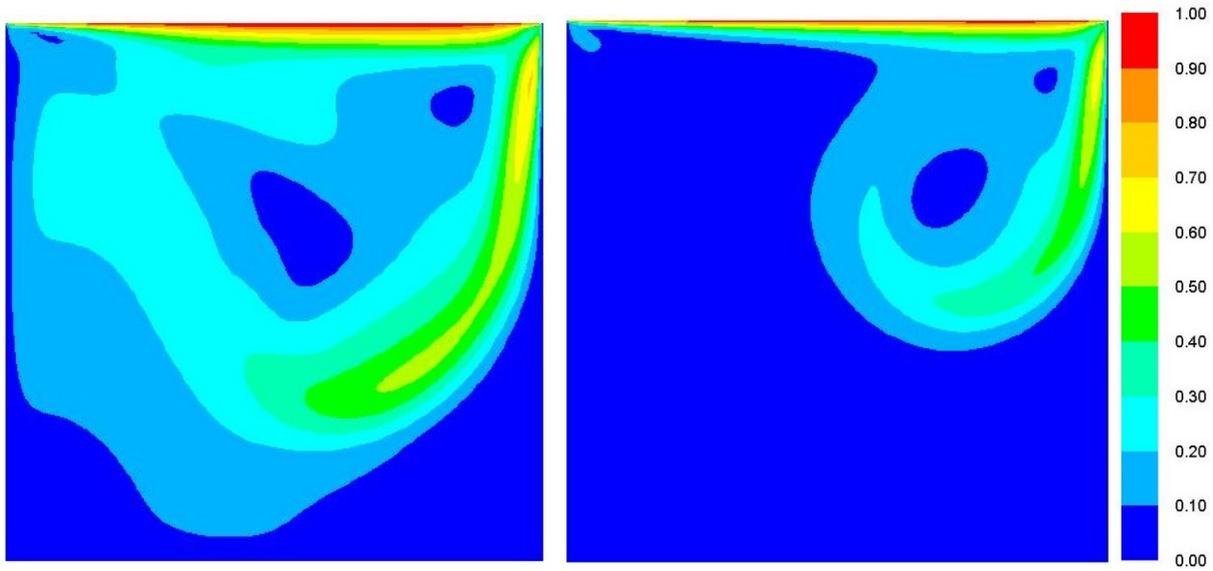

(c)                  (d)

Figure 15: Mixing in Case A with density ratio 2:1, developing flow at T=5: (a) mass fraction of the first fluid, (b) isolines of $\nabla \cdot \vec{V}$, (c) velocity magnitude, and (d) velocity magnitude for Case A with uniform density at T=5, presented for comparison. Concentration field in sub-figure (a) has the same bar as in Fig. 5.

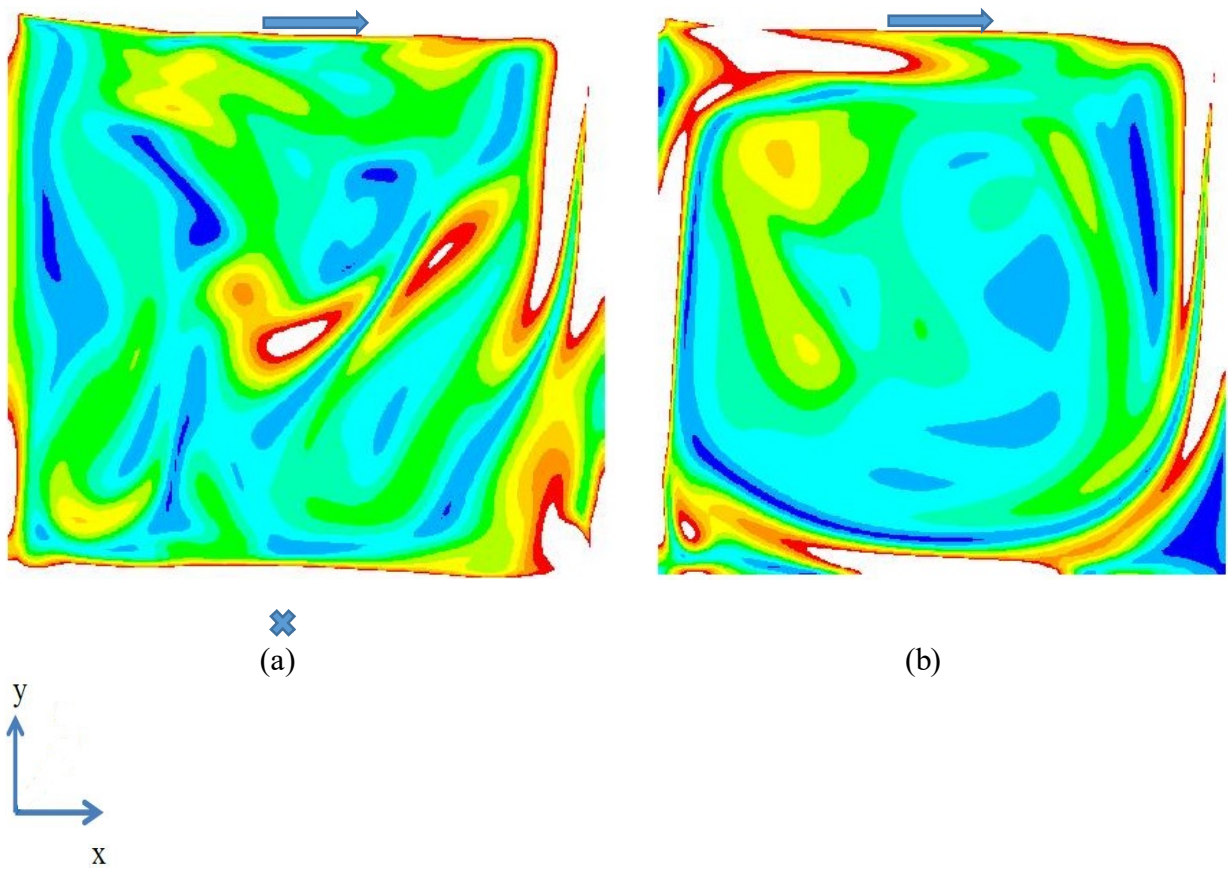

Figure 16: Vorticity distribution in developed flow with density ratio 2:1: (a) Case A and (b) Case C

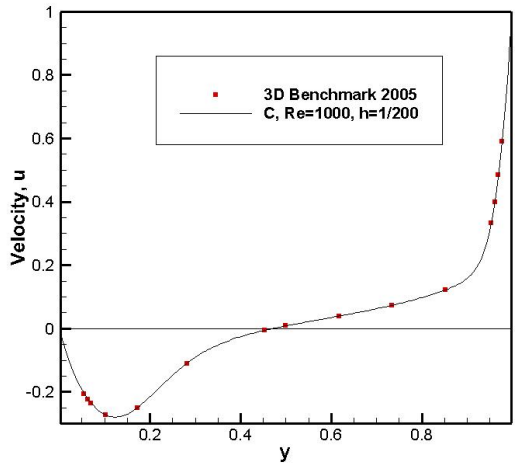 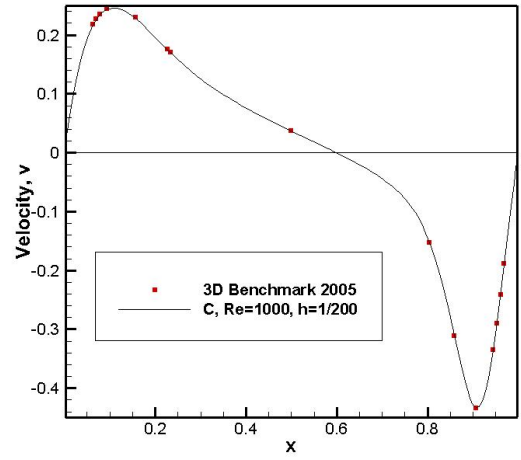

(a)                      (b)

Figure A-1: Comparison of current computations (Case C) with benchmark[18]: a) velocity component, u, along vertical centerline, (0.5, 0≤y≤1, 0.5) and b) velocity component, v, along horizontal centerline, (0≤x≤1, 0.5, 0.5).